\documentclass[aps,prx,reprint,groupedaddress,showpacs,longbibliography]{revtex4-1} 
\usepackage{amssymb,amsmath}
\usepackage{graphicx}
\usepackage{xcolor}
\usepackage[latin1]{inputenc}
\usepackage[T1]{fontenc}

\usepackage{siunitx}

\newcommand{\bs}{\boldsymbol}
\newcommand{\D}{{\rm d}}
\newcommand{\I}{{\rm i}}
\newcommand{\E}{{\rm e}}

\begin{document}

\title{Dynamical symmetry and breathers in a two-dimensional Bose gas}

\author{R. Saint-Jalm, P.C.M. Castilho, \'E. Le Cerf, B. Bakkali-Hassani, J.-L. Ville, S. Nascimbene, J. Beugnon, J. Dalibard}

\email[]{jean.dalibard@lkb.ens.fr}
 
\affiliation{Laboratoire Kastler Brossel,  Coll\`ege de France, CNRS, ENS-PSL University, Sorbonne Universit\'e, 11 Place Marcelin Berthelot, 75005 Paris, France}

\date{\today}
\begin{abstract}
\noindent
A fluid is said to be \emph{scale-invariant} when its interaction and kinetic energies have the same scaling in a dilation operation. In association with the more general conformal invariance, scale invariance provides a dynamical symmetry which has profound consequences both on the equilibrium properties of the fluid and its time evolution. Here we investigate experimentally the far-from-equilibrium dynamics of a cold two-dimensional rubidium Bose gas. We operate in the regime where the gas is accurately described by a classical field obeying the Gross--Pitaevskii equation, and thus possesses a dynamical symmetry described by the Lorentz group SO(2,1). With the further simplification provided by superfluid hydrodynamics, we show how to relate the evolutions observed for different initial sizes, atom numbers, trap frequencies and interaction parameters by a scaling transform. Finally we show that some specific initial shapes - uniformly-filled  triangles or disks - may lead to a periodic evolution, corresponding to a novel type of breather for the two-dimensional Gross--Pitaevskii equation.
\end{abstract}

\maketitle


\section{Introduction}
\label{sec:introduction}

Symmetries play a central role in the investigation of a physical system.  Most often they are at the origin of conserved quantities, which considerably simplify the study of the  equilibrium states and the evolution of the system. For example, spatial symmetries associated with translation and rotation lead to the conservation of linear and angular momentum. More generally, it is interesting to determine the dynamical (or hidden) symmetries of the system under study, which can lead to more subtle conserved quantities. These symmetries are described by the group of all transformations of space and time that leave the action, therefore the equations of motion, invariant. A celebrated example is the $1/r$ potential in three dimensions, where there exists a dynamical symmetry described by the group O(4) for the bounded orbits \cite{Bander:1966}. When treated by classical mechanics, it leads to the conservation of the Laplace--Runge--Lenz vector, from which one deduces that the bounded orbits are actually closed trajectories.

Among the systems that display rich dynamical symmetries are the ones whose action is left invariant by a dilation transformation of space and time. Such scale-invariant systems were initially introduced in particle physics to explain scaling laws in high energy collisions \cite{Jackiw:1972}. We will consider here the non-relativistic version of scale-invariance, which applies to the dynamics of a fluid of particles. We consider the simultaneous change of length and time coordinates of each particle according to the scaling:
\begin{equation}
\bs r\to  \bs r/\lambda,\qquad t \to  t/\lambda^2.
\label{eq:scaling}
\end{equation}
In this dilation, the velocity of a particle is changed as $\bs v\to \lambda \bs v$. Therefore the kinetic energy of the fluid scales as $E_{\rm k}\to \lambda^2E_{\rm k}$, which ensures that the corresponding part of the action ($\propto \int E_{\rm k}\; \D t$) remains invariant in the transformation (\ref{eq:scaling}). If the interaction energy has the same scaling,  $E_{\rm i}\to \lambda^2E_{\rm i}$, the total action of the fluid is invariant in the dilation. The simplest example of such a fluid is a collection of non-relativistic particles, either non-interacting ($E_{\rm i}=0$) or with pair-wise interactions described by a $1/r^2$ potential. A scale-invariant fluid  possesses remarkably simple thermodynamic properties: for example its equation of state depends only on the ratio of chemical potential to temperature, instead of being an independent function of these two variables. 

Most physical systems exhibiting scale invariance also possess a more general conformal invariance, where time and space are modified by conformal transformations instead of the simple dilations given in Eq.(\ref{eq:scaling}) \cite{Nakayama:2013}. 
In the non-relativistic domain,  this conformal invariance exists for the Schr\"odinger equation describing the motion of the two systems mentioned above, free particles \cite{Hagen:1972,Niederer:1972} and particles interacting with a $1/r^2$ potential \cite{Alfaro:1976}. In both cases, the dynamical symmetry group 
associated to this scale/conformal invariance is the Lorentz group SO(2,1). This is also the case for the three-dimensional pseudo-spin\,$1/2$ Fermi gas in the unitary regime (for a review, see e.g. \cite{Zwerger:2011}). There, the scattering length between the two components diverges, ensuring the required disappearance of a length scale related to interactions. In addition to the existence of a universal equation of state, this dynamical symmetry leads to a vanishing bulk viscosity \cite{Son:2007,Elliott:2014}, and also to general relations between the moments of the total energy and those of the trapping energy in a harmonic potential \cite{Werner:2006}.

In this article we consider another example of a scale/conformal invariant fluid with the SO(2,1) dynamical symmetry, the "weakly interacting" two-dimensional (2D) Bose gas. The concept of "weak interaction" means in this context that the state of the gas is well described by a classical field $\psi(\bs r,t)$. This field is normalized to unity ($\int |\psi|^2\;\D^2 r=1$), so that the density of the gas reads $n(\bs r,t)=N|\psi(\bs r,t)|^2$ where $N$ is the number of particles. In the scaling of positions, the 2D matter-wave field changes as
$\psi(\bs r) \to \lambda\, \psi(\lambda\bs r)$, which guarantees that the norm is preserved and that the dynamical part of the action $\propto \I \hbar \int \D t\int \D^2 r\; \psi^*\,\partial_t \psi$ is invariant. The interaction energy of the gas then reads for contact interaction
\begin{equation}
E_{\rm i}=\frac{N^2\hbar^2}{2m}\tilde g  \int |\psi(\bs r)|^4\;\D^2 r,
\label{eq:interaction_2D}
\end{equation}
where $m$ is the mass of a particle and $\tilde g$ the dimensionless parameter characterizing the  strength of the interaction. One can immediately check that $E_{\rm i}$ obeys the $\lambda^2$ scaling required for scale invariance, which can be viewed as a consequence of the dimensionless character of $\tilde g$. The classical field description used here is valid if one restricts to the case of a small coupling strength $\tilde g \ll 1$ \cite{Svistunov:2015}.  This restriction is necessary because of the singularity of the contact interaction $\frac{\hbar^2}{m}\tilde g \delta(\bs r)$ in 2D when it is treated at the level of quantum field theory. Note that the condition $\tilde g\ll 1$ does not constrain the relative values of the interaction and kinetic energies. Actually in the following we will often consider situations where $E_{\rm i}\gg E_{\rm k}$ (Thomas--Fermi regime). 

So far, the scale/conformal invariance of the weakly interacting 2D Bose gas has been mainly exploited to measure its equation of state \cite{Hung:2011,Yefsah:2011}. 
Also, one of its dynamical consequences in an isotropic 2D harmonic potential of frequency $\omega$ has been explored: the frequency of the breathing mode was predicted to be exactly equal to $2\omega$ for any $\tilde g$ \cite{Kagan:1996b,Pitaevskii:1997a,gritsev:2010}, as tested in Refs.\, \cite{Chevy:2002,Vogt:2012}. Note that in the presence of a harmonic potential, the whole system is not scale-invariant anymore, but it still possesses a dynamical symmetry described by the group SO(2,1), as shown in \cite{Pitaevskii:1997a}. Recently, deviations from this prediction for $\tilde g \gtrsim 1$, an example of a \emph{quantum anomaly} \cite{Olshanii:2010}, have been observed \cite{Peppler:2018,Holten:2018}.

The purpose of our work is to go beyond static properties of the weakly interacting 2D Bose gas and its single-mode oscillation in a harmonic potential, and to reveal  more general features associated with its dynamical symmetry. To do so, we study the evolution of the gas in a 2D harmonic potential of frequency $\omega$, starting from a uniformly-filled simple area (disk, triangle or square). Here we use $\tilde g\leq 0.16$ so that the classical field description is legitimate. We first check (Sec.\,\ref{sec:periodic_Ep}) the prediction from \cite{Pitaevskii:1997a} that $E_{\rm k}+E_{\rm i}$ should have a periodic evolution in the trap with the frequency $2\omega$. We then investigate the transformations linking different solutions of the equations of motion. These transformations are at the heart of the dynamical symmetry group SO(2,1). In practice we first link the evolution of clouds with the same atom number and homothetic initial wave functions in harmonic potentials with different frequencies (Sec.\,\ref{sec:same_Nat}). Then, restricting to the case where superfluid hydrodynamics is valid, we derive and test a larger family of transformations that allows one to connect the evolutions of two initial clouds of similar shapes with different sizes, atoms numbers, trap frequencies and interaction strengths (Sec.\,\ref{sec:vary_gN}). Finally in Sec.\,\ref{sec:breathers} we explore a property that goes beyond the symmetry group of the system and that is specific to triangular and disk-shaped distributions in the hydrodynamic limit: we find numerically that these distributions evolve in a periodic manner in the harmonic trap, and we confirm this prediction over the accessible range for our experiment (typically two full periods of the trap $4\pi/\omega$). These particular shapes can therefore be viewed as two-dimensional breathers for the Gross--Pitaevskii (non-linear Schr\"odinger) equation in the hydrodynamic limit \cite{dauxois2006physics}. They also constitute a novel example of universal dynamics in a quantum system prepared far from equilibrium \cite{Eigen:2018aa,Ern:2018,Prufer:2018aa}. 


\section{Evolution of potential energy}
\label{sec:periodic_Ep}

Our experiment starts with a 3D Bose-Einstein condensate of $^{87}$Rb  that we load around a single node of a vertical ($z$) standing wave created with a laser of wavelength 532\,nm.  The confining potential  along $z$ is approximately harmonic with a frequency $\omega_z/(2\pi)$ up to 4.9\,kHz. The interaction parameter is $\tilde g=\sqrt{8\pi}\,a_s/\ell_z$, where $a_s$ is the 3D s-wave scattering length and $\ell_z=(\hbar/m\omega_z)^{1/2}$. The interaction energy per particle and the residual temperature are both smaller than $\hbar \omega_z$ so that the vertical degree of freedom is effectively frozen \cite{Ville:2017}. The initial confinement in the horizontal $xy$ plane is ensured by "hard walls" made with a light beam also at 532\,nm. This beam is shaped using a digital micromirror device (DMD), and a high-resolution optical system images the DMD pattern onto the atomic plane \cite{Aidelsburger:2017b}, creating a box potential on the atoms. The cloud fills uniformly this box potential, and it is evaporatively cooled by adjusting the height of the walls of the box. For all data presented here, we kept the temperature low enough to operate deep in the superfluid regime, $T/T_c<0.3$, where $T_c$ is the critical temperature for the Berezinskii-Kosterlitz-Thouless transition. At this stage the atoms are prepared in the $F=1,m_F=0$ hyperfine (ground) state, which is insensitive to magnetic field.

Once the gas has reached equilibrium in the 2D box, we suddenly switch off the confinement in the $xy$ plane and simultaneously transfer the atoms to the field-sensitive state $F=1,m_F=-1$ using two consecutive microwave transitions, via the intermediate state $F=2,m_F=0$. Most of the experiments are performed in the presence of a magnetic field that provides the internal state $F=1,m_F=-1$ with an isotropic harmonic confinement in the $xy$ plane, with $\omega/2\pi$ around $19.5$\,Hz. We estimate the anisotropy of the potential to be $\lesssim 2\%$. We let the cloud evolve in the harmonic potential for an adjustable time before making an in-situ measurement of the spatial density $n(\bs r)=N|\psi(\bs r)|^2$ by absorption imaging.

The measurement of $n(\bs r)$ gives access to both the interaction energy (\ref{eq:interaction_2D}) and the potential energy in the harmonic trap 
\begin{equation}
E_{\rm p}=\frac{N}{2}m\omega^2 \int r^2\,|\psi(\bs r)|^2\;\D^2 r.
\end{equation}
Since the gas is an isolated system, we expect the total energy $E_{\rm tot}=E_{\rm k}+E_{\rm i}+E_{\rm p}$ to be conserved during the evolution, where the kinetic energy $E_{\rm k}$ reads:
\begin{equation}
E_{\rm k}=\frac{N\hbar^2}{2m}\int |\bs \nabla \psi|^2\;\D^2 r.
\end{equation}
The SO(2,1) symmetry for a 2D harmonically trapped gas brings a remarkable result: $E_{\rm k}+E_{\rm i}$ and $E_{\rm p}$ should oscillate sinusoidally at frequency $2\omega$ \cite{Pitaevskii:1997a}. More precisely, using the 2D Gross--Pitaevskii equation one obtains the relations
\begin{eqnarray}
\frac{\D E_{\rm p}}{\D t}&=&-\frac{\D (E_{\rm k}+E_{\rm i})}{\D t}=\omega W, \label{eq:evolution}\\ 
\frac{\D W}{\D t}&=&2\omega(E_{\rm k}+E_{\rm i}-E_{\rm p}),
\label{eq:evolutionb}
\end{eqnarray} 
where we have defined $W=\omega m \int \bs r\cdot\bs v\,n\ \D^2\bs r$, and the velocity field $\bs{v}(\bs r) = \frac{\hbar}{m}\mbox{Im}\left[\psi^*(\bs r) \bs \nabla \psi(\bs r)\right]/|\psi(\bs r)|^2$.
Initially the gas is prepared in a steady state in the box potential, so that $\bs v=0$ hence $W(0)$ is also null. Therefore the potential energy evolves as
\begin{equation}
E_{\rm p}(t)=\frac{1}{2}E_{\rm tot} + \Delta E\, \cos(2\omega t),
\label{eq:Epot_vs_t}
\end{equation}
where $\Delta E=\frac{1}{2}[E_{\rm p}(0)-E_{\rm k}(0)-E_{\rm i}(0)]$ can be positive or negative. A similar result holds for the sum $E_{\rm k}+E_{\rm i}$ (with $\Delta E$ replaced by $-\Delta E$), but not for the individual energies $E_{\rm k}$ or $E_{\rm i}$.

\begin{figure}
\begin{center}
\includegraphics[width=\columnwidth]{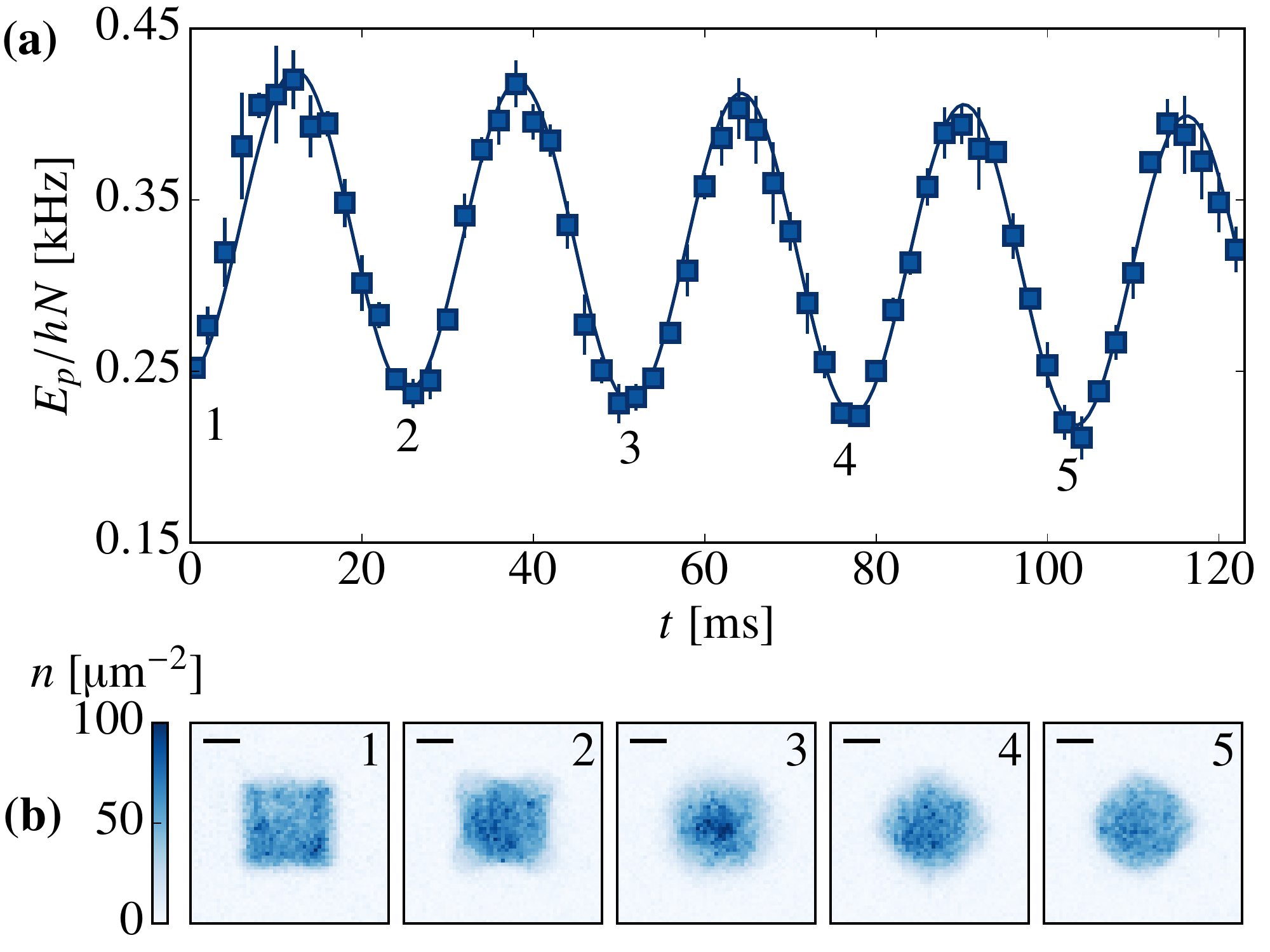}
\end{center}
\caption{Time evolution of the potential energy per particle of a 2D gas of $^{87}$Rb atoms in an isotropic harmonic potential of frequency $\omega$ for a square of side length $27.6(5)\,\si{\micro\meter}$ with $4.1(2)\times10^4$ atoms. (a): Evolution of the potential energy per particle. Each point is an average of 7 to 10 realizations, and the error bars indicate the standard deviation of these different realizations. The frequency of the trap is measured with the oscillation of the center of mass: $\omega/2\pi= 19.3(1)\ \si{\hertz}$. The oscillations of $E_{\rm p}$ are fitted with a cosine function and an additional linear slope (continuous line). This slope is $-0.25(4)\ \si{\hertz/\milli\second}$ and accounts for the loss of particles from the trap. The fitted frequency is $ 38.5(1)\ \si{\hertz}$, which is compatible with $\omega/\pi$, as predicted by the SO(2,1) symmetry of the gas. (b): Density distribution of an initially uniform gas after the evolution in a harmonic potential at times $\omega t = 0,\,\pi,\,2\pi,\,3\pi,\,4\pi$, corresponding to the first periods of the potential energy indicated by the labels from 1 to 5. The horizontal black lines represent $10\,\si{\micro\meter}$.}
\label{fig:periodic_evol_Ep}
\end{figure}

We show in Fig.\,\ref{fig:periodic_evol_Ep}(a) the evolution of the potential energy per particle for an initially uniformly-filled square. Although the density distribution is not periodic (see Fig.\,\ref{fig:periodic_evol_Ep}(b)), the potential energy $E_{\rm p}$ evolves periodically and is well fitted by a cosine function with a period that matches the $2\omega$ prediction and the expected zero initial phase. For a better adjustment of the data, we added a (small) negative linear function to the fitting cosine. Its role is likely to account for the residual evaporation rate of atoms from the trap ($\sim 0.1\ \si{\second}^{-1}$).

This simple dynamics can be viewed as a generalisation of the existence of the undamped breathing mode at frequency $2\omega$ that we mentioned in the introduction \cite{Kagan:1996b,Pitaevskii:1997a}. We emphasize that this result is a consequence of the SO(2,1) symmetry and would not hold for the Gross--Pitaevskii equation in 1D or 3D. 

\newcommand{\ta}{t}
\newcommand{\tb}{t'}
\newcommand{\ra}{\bs r}
\newcommand{\rb}{\bs {r}'}
\newcommand{\rbn}{{r}'}
\newcommand{\pL}{\alpha}

\section{General scaling laws\label{sec:same_Nat}}

An important consequence of the dynamical symmetry of the 2D Gross--Pitaevskii equation is the ability to link two solutions $\psi_{1,2}$ of this equation corresponding to homothetic initial conditions: one can relate $\psi_1(\bs r,t)$ and $\psi_2(\bs r',t')$, provided they evolve with the same parameter $\tilde g N$ and the same trap frequency $\omega_1= \omega_2$. By using a simple scaling on space and time, this link can be further extended to the case $\omega_1\neq \omega_2$.  

The general procedure is presented in Appendix \ref{app:SO(2,1)} and we start this section by summarizing the main results. Consider a solution of the Gross--Pitaevskii equation $\psi_1(\bs r,t)$ for the harmonic potential of frequency $\omega_1$:
\begin{equation}
\I \hbar \frac{\partial \psi_1}{\partial \ta}=-\frac{\hbar^2}{2m}\bs \nabla^2\psi_1+\frac{\hbar^2\tilde gN}{m}|\psi_1|^2 \psi_1 +\frac{1}{2}m\omega_1^2 \ra^2 \psi_1. 
\label{eq:GPeq}
\end{equation}
Using scale/conformal invariance, we can construct a solution $\psi_2(\bs r',t')$ of the Gross--Pitaevskii equation with the frequency $\omega_2=\zeta \omega_1$ using:
\begin{equation}
\psi_2(\rb,\tb)=f(\ra,\ta)\,\psi_1(\ra,\ta)
\label{eq:transform_lambda2a}
\end{equation}
where space is rescaled by $\rb={\ra}/{\lambda(\ta)}$ with
\begin{equation}
\lambda(\ta)= \left[\frac{1}{\pL^2}\cos^2(\omega_1 \ta)+\pL^2\zeta^2 \sin^2 (\omega_1 \ta)   \right]^{1/2},
\label{eq:lambda_t}
\end{equation}
where the dimensionless parameter $\pL$ is the homothetic ratio between the initial states. The relation between the times $\ta$ and $\tb$ in frames 1 and 2 is
\begin{equation}
\tan(\omega_2 \tb)=\zeta\pL^2 \, \tan(\omega_1 \ta),
\label{eq:relation_t_tau}
\end{equation}
and the multiplicative function $f$ is
\begin{equation}
f(\ra,\ta)=\lambda(\ta) \exp\left(-\I \frac{m\dot \lambda \ra^2}{2\hbar \lambda}  \right),
\end{equation}
where $\dot \lambda\equiv\frac{\D \lambda}{\D \ta}$. 
The two solutions $\psi_{1,2}(t)$ correspond to the evolution of two clouds with the same parameter $\tilde g_1 N_1=\tilde g_2 N_2$. At $t = 0$, these two wave functions correspond to the ground states of the Gross-Pitaevskii equation in the box potentials with characteristic lengths $L_{1,2}$, with $L_2=\pL L_1$. Both initial wave functions $\psi_{1,2}(0)$ can be chosen real, and the scale invariance of the (time-independent) 2D Gross-Pitaevskii equation ensures that they are homothetic: $\pL \psi_2(\pL \bs r,0)= \psi_1( \bs r,0)$. For example in the limit $E_{\rm i}\gg E_{\rm k}$, $\psi(0)$ corresponds to a uniform density in the bulk, and goes to zero at the edges on a scale given by the healing length $\xi\equiv [N\hbar^2/(2mE_{\rm i})]^{1/2}$. For two box potentials of homothetic shapes filled with the same number of particles, the ratio $\xi_2/\xi_1$ is equal to the ratio $L_2/L_1$.

\begin{figure*}[t]
\begin{center}
\includegraphics[width=2\columnwidth]{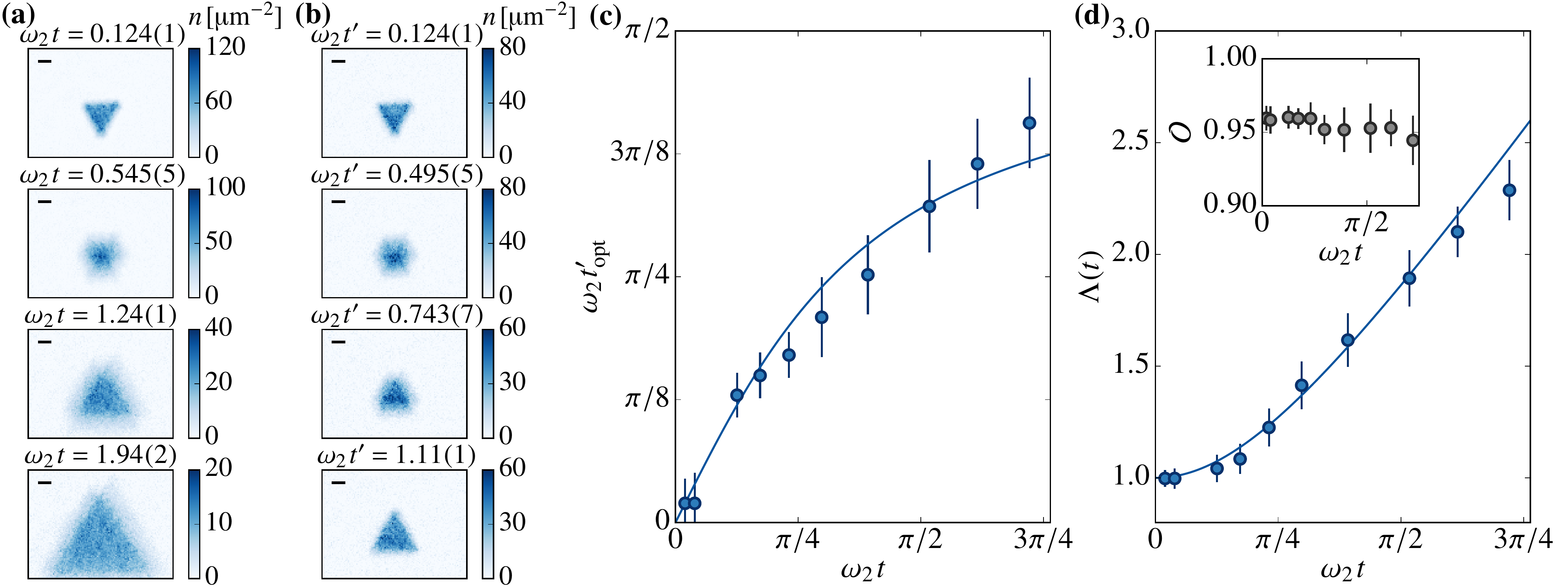}
\end{center}
\caption{Evolution of a gas with triangular shape (side length $40.2(3)\ \si{\micro\meter}$, $3.9(3)\times 10^4$ atoms) for two different values of the harmonic trapping frequency. (a) and (b): Averaged images of the density distribution after a variable evolution time in the harmonic potential of frequency $\omega_1 = 0$ and $\omega_2/2\pi = 19.7(2)\ \si{\hertz}$ respectively. The images result of an average over 5 to 10 realizations, and the horizontal black lines represent $10\,\si{\micro\meter}$. Pairs of images with approximately corresponding evolution times have been chosen. (c): Optimal time $\tb_{\rm opt}(\ta)$ for which the overlap between images of the first and the second evolutions is maximum. (d): Optimal rescaling factor between the corresponding images $n_1(\ta)$ and $n_2(\tb_{\rm opt})$. On the two graphs (c) and (d), the solid lines are the theoretical predictions given by (\ref{eq:tau_part}) and (\ref{eq:lambda_part}). The inset of figure (d) presents the overlap between the corresponding images of the two series. On (c) and (d), the error bars indicate the confidence intervals within two standard deviations of the fits used to reconstruct the scaling laws.}
\label{fig:TOF_vs_osc}
\end{figure*}

We explore experimentally this mapping between two evolutions in the particular case $L_1=L_2$  and $\omega_1 \to 0$, i.e., $\pL=1$ and $\zeta\to +\infty$. This corresponds to comparing the evolution of clouds with the same shape and the same size either in a harmonic potential or in free (2D) space. The choice of the initial shape is arbitrary; here we start from a uniform triangle of side length $40.2(3)\ \si{\micro\meter}$ with $3.9(3)\times 10^4$ atoms and let it evolve either in a harmonic potential of frequency $\omega_2/(2\pi) = 19.7(2)\,\si{\hertz}$, or without any potential ($\omega_1 = 0$). In both cases, we record images of the evolution, examples of which are given in Fig.\,\ref{fig:TOF_vs_osc}(a) and \ref{fig:TOF_vs_osc}(b). These two evolutions should be linked via Eq.\,(\ref{eq:transform_lambda2a}). The relation (\ref{eq:relation_t_tau}) between $\ta$ and $\tb$ reads
\begin{equation}\label{eq:tau_part}
\tan(\omega_2 \tb)=\omega_2 \ta,
\end{equation}
and the relation (\ref{eq:lambda_t}) becomes:
\begin{equation}\label{eq:lambda_part}
\lambda(\ta) = (1+\omega_2^2\ta^2)^{1/2}.
\end{equation}
The relation (\ref{eq:tau_part}) indicates that the scaling transformation maps the first quarter of the oscillation period in the harmonic trap $\omega_2\tb \leq \pi/2$ onto the ballistic expansion from $\ta=0$ to $\ta=\infty$. In the absence of interactions, this result has a simple physical interpretation: after the ballistic expansion between $\ta=0$ to $\ta=\infty$, the asymptotic position distribution reveals the initial velocity distribution of the gas, whereas the evolution in the harmonic trap during a quarter of oscillation period exchanges initial positions and initial velocities. 
We emphasize that the mapping (\ref{eq:tau_part}) also holds for an interacting system, as a consequence of the SO(2,1) symmetry underlying the Gross--Pitaevskii equation 
\footnote{ One may question the validity of the Gross-Pitaevskii equation (GPE), hence of scale invariance, after a long expansion time when the gas occupies a large area $R^2$. If we were interested in the ground state of a box of size $R\to \infty$ and a given $\tilde gN$, we would indeed expect deviations with respect to GPE because the relevant momenta $k\sim R^{-1}$ would tend to 0, and logarithmic corrections in $k$ to the coupling constant would become significant \cite{Petrov:2000a}. Here, this issue is absent because the initial interaction energy is converted into kinetic energy at the beginning of the expansion. The relevant atomic momenta thus remain $\sim (2mE_{\rm int}/N\hbar^2)^{1/2}$ at all times, which validates the use of the GPE. }.

In order to reconstruct the scaling laws (\ref{eq:tau_part}) and (\ref{eq:lambda_part}) from the measured evolutions, we compare each image $n_1(\ra,\ta)$ for the free evolution with the set of images $n_2(\rb,\tb)$ obtained for the in-trap evolution. More precisely, we start by defining the overlap ${\cal O}[n_1,n_2]$ between two images in the following way: 
\begin{itemize}
 \item
We introduce the scalar product $(n_1|n_2)$ between two images
\begin{equation}
(n_1|n_2)=\int n_1(\bs r)\,n_2(\bs r)\;d^2 r
\end{equation}
and the norm of an image $||n_1||=\sqrt{(n_1|n_1)}$.

 \item
In order to relate two images that differ by a spatial scaling factor $\lambda$, we introduce the quantity
\begin{equation}
p[n_1,n_2,\lambda]=\frac{(n_1^{(\lambda)}|n_2)}{||n_1^{(\lambda)}||\;||n_2||}
\label{eq:def_Pn1n2}
\end{equation}
where $n_1^{(\lambda)}(\bs r)=\lambda^2 n_1(\lambda \bs r)$ is the image rescaled from $n_1(\bs r)$ by the factor $\lambda$, with the same atom number: $N_1=\int n_1(\bs r)\;\D^2 r=   \int n_1^{(\lambda)}(\bs r)\;\D^2 r$. Note that the definition of the norm given above entails $||n_1^{(\lambda)}||=\lambda||n_1||$. By construction the quantity $p[n_1,n_2,\lambda]$ is always smaller or equal to 1, and it is equal to 1 only when the image $n_1^{(\lambda)}$ is identical to $n_2$ up to a multiplicative factor. 

\item
Finally, for a couple of images $(n_1,n_2)$ we vary $\lambda$ and define their overlap as
\begin{equation}
{\cal O}[n_1,n_2]=\max_\lambda p[n_1,n_2,\lambda].
\label{eq:definition_overlap}
\end{equation}

\end{itemize}
In practice, for each image $n_1(\ta)$ acquired at a given time $\ta$, we determine the time $\tb_{\rm opt}$ where the overlap between $n_1(\ta)$ and $n_2(\tb)$ is optimal. We  denote $\Lambda(t)$ the value of the scaling parameter $\lambda$ for which the value ${\cal O}[n_1(\ta),n_2(\tb_{\rm opt})]$ is reached (see Supplemental material for more details). Since the center of the images may drift during the evolution, we also allow for a translation of $n_2$ with respect to $n_1$ when looking for the optimum in (\ref{eq:def_Pn1n2}-\ref{eq:definition_overlap}).

The result of this mapping between the two evolutions is shown in Fig.\,\ref{fig:TOF_vs_osc}(c) and \ref{fig:TOF_vs_osc}(d). In Fig.\,\ref{fig:TOF_vs_osc}(c), we plot $\tb_{\rm opt}$ as a function of $\ta$. The prediction (\ref{eq:tau_part}) is shown as a continuous line and is in good agreement with the data. In Fig.\,\ref{fig:TOF_vs_osc}(d), we show the variation of the corresponding optimal scaling parameter $\Lambda(t)$. Here again the prediction (\ref{eq:lambda_part}) drawn as a continuous line is in good agreement with the data. The overlap between the density distributions at the corresponding times is shown in the inset of Fig.\,\ref{fig:TOF_vs_osc}(d) and is always around $0.95$, confirming that these density distributions have very similar shapes. Indeed, the overlap between two images averaged over a few experimental realizations, taken in the same conditions ranges from $0.98$ to $0.99$ due to experimental imperfections.

Finally we note that here we connected solutions of the Gross-Pitaevskii equation (\ref{eq:GPeq}) with the same atom number $N_1=N_2$. Actually the results derived above also apply to pairs of solutions with $\tilde g_1N_1=\tilde g_2N_2$, since only the product $\tilde g N$ enters in the Gross-Pitaevski equation (\ref{eq:GPeq}). 


\section{Scaling laws in the hydrodynamic regime\label{sec:vary_gN}}

In the previous section, we have linked the evolution of two clouds with the same atom number $N$ (or the same $\tilde{g}N$). We show now that it is also possible to link evolutions with different $N$'s and $\tilde{g}$'s, provided we restrict to the so-called hydrodynamic (or Thomas--Fermi) regime, where the healing length $\xi$ is very small compared to the size of the gas. 


\subsection{General formulation}

The Gross--Pitaevskii equation (\ref{eq:GPeq}) can be equivalently written in terms of the density and the velocity fields as 
\begin{gather}
\partial_t n + \bs{\nabla}\cdot(n\bs{v}) = 0,\label{eq:continuity2}\\
m\partial_t\bs{v} + \mathbf{\bs \nabla} \left(\frac{1}{2}m\bs{v}^2 + \frac{\hbar^2}{m}\tilde{g}n + \frac{1}{2}m\omega^2 r^2 + P(n)\right) = 0,\label{eq:GPE2}
\end{gather}
where $P(n) = -\hbar^2/2m\ (\bs \nabla^2\sqrt{n})/\sqrt{n}$ is the so-called quantum pressure. When the characteristic length scales over which the density and velocity vary are much larger than the healing length $\xi$, one can neglect the contribution of the quantum pressure in (\ref{eq:GPE2}): 
\begin{equation}
m\partial_t\bs{v} + \mathbf{\bs \nabla} \left(\frac{1}{2}m\bs{v}^2 + \frac{\hbar^2}{m}\tilde{g}n + \frac{1}{2}m\omega^2 r^2\right) = 0.\label{eq:GPE2_approx}
\end{equation}
This approximation, corresponding to the Thomas--Fermi limit, leads to the regime of quantum hydrodynamics for the evolution of the density $n$ and the irrotational velocity field $\bs v$ \cite{Pitaevskii:2016}. It enriches the dynamical symmetries of the problem, as we see in the following. For our experimental parameters, this approximation is legitimate since the healing length is a fraction of micrometer only, much smaller than the characteristic size of our clouds (tens of micrometers). 

We consider two homothetic shapes, e.g. two box-like potentials with a square shape, with sizes $L_{1,2}$ and filled with $N_{1,2}$ atoms. We assume that we start in both cases with the ground state of the cloud in the corresponding shape, so that the initial velocity fields are zero. Note that contrarily to the case of Sec.\,\ref{sec:same_Nat}, the ratio between the healing lengths $\xi_2/\xi_1$ is not anymore equal to $L_2/L_1$ so that the initial wave functions are not exactly homothetic, but this mismatch occurs only close to the edges over the scale $\sim \xi_{1,2} \ll L_{1,2}$. As before, at time $t=0$ we switch off the potential creating the shape under study, and switch on a harmonic potential with frequency $\omega_{1,2}$. Our goal is to relate the two evolutions with parameters $(\tilde{g}_1N_1,L_1,\omega_1)$ and $(\tilde{g}_2N_2,L_2,\omega_2)$.

The general transformation involves three dimensionless constant parameters $\mu, \pL, \zeta$:
\begin{equation}
\tilde{g}_2 N_2=\mu^2\, \tilde{g}_1N_1,\quad L_2=\pL\,L_1,\quad \omega_2=\zeta\,\omega_1,
\label{eq:scaling_general}
\end{equation}
 and reads:
\begin{eqnarray}
\tilde{g}_2 n_2(\rb,\tb) &=& \lambda^2\mu^2\,\tilde{g}_1 n_1(\ra,\ta), \label{eq:transform_lambda3a} \\
\bs v_2(\rb,\tb)&=& \lambda\mu\, \bs v_1(\ra,\ta) -\mu \dot\lambda\ra.
\label{eq:transform_lambda3b}
\end{eqnarray}
with $\dot\lambda=\frac{\D \lambda}{\D \ta}$.
The spatial variables are rescaled as $\rb =\ra/\lambda(\ta)$ with the function $\lambda$ now given by
\begin{equation}
\lambda(\ta)=\left[\frac{1}{\pL^2}\cos^2(\omega_1 \ta)+\left( \frac{\zeta\pL}{\mu} \right)^2\sin^2(\omega_1 \ta)  \right]^{1/2},
\label{eq:lambda_t_bis}
\end{equation}
and the relation between the times $\ta$ and $\tb$ in frames 1 and 2 is:
\begin{equation}
\tan(\omega_2 \tb)= \frac{\zeta\pL^2}{\mu} \tan(\omega_1 \ta).
\label{eq:relation_t_tau_bis}
\end{equation}
With a calculation similar to that detailed in Appendix \ref{app:SO(2,1)}, one can readily show that if $(n_1,\bs v_1)$ is a solution of the hydrodynamic equations (\ref{eq:continuity2},\ref{eq:GPE2_approx}) for the frequency $\omega_1$, then $(n_2,\bs v_2)$ is a solution for the frequency $\omega_2$. 
If $\mu = 1$, these equations also apply beyond the Thomas--Fermi limit, as shown in Sec.\,\ref{sec:same_Nat}.
More strikingly, they show that, in the quantum hydrodynamic regime,  the evolution of any cloud is captured by a universal dynamics that depends only on its initial geometry.


\subsection{Connecting evolutions with a fixed trap frequency, a fixed size and different $\tilde{g} N$}

We present here the experimental investigation of the scaling described above, focusing on the case $L_1=L_2$ and $\omega_1=\omega_2$, {i.e.}, $\pL=\zeta=1$. In other words, we compare the evolution of two clouds with the same initial shape and density distribution, different atom numbers and different interaction strengths in a given harmonic trap. For simplicity, we consider the result of the evolution at times $\ta$ and $\tb$ such that $\omega_1\ta=\omega_2 \tb=\pi/2$, which satisfies the constraint (\ref{eq:relation_t_tau_bis}). In this case $\lambda(\ta)=1/\mu$, so that the general scaling (\ref{eq:transform_lambda3a}) reads
\begin{equation}
\tilde g_2 n_2(\mu\ra,\tb_{\pi/2})=\tilde g_1 n_1\left(\ra,\ta_{\pi/2}\right).
\label{eq:scaling_gN}
\end{equation}

We start with a cloud in a uniform box potential with the shape of an equilateral triangle of side length $L = 38.2(3)\ \si{\micro\meter}$. At $t = 0$ we transfer the atoms in the harmonic trap of frequency $\omega/2\pi= 19.6\ \si{\hertz}$ and remove the box potential. At $t = \pi/(2\omega)$ we image the cloud. We perform this experiment for different values of $\tilde{g}$ -- and slightly different atom numbers -- corresponding to the product $\tilde gN$ between 200 and $4000$. This leads to a ratio $\xi/L$ always smaller than $0.03$, ensuring that we stay in the quantum hydrodynamic regime. The variation of $\tilde g$ is achieved by changing the intensity $I$ of the laser beams creating the vertical confinement, with $\tilde g\propto I^{1/4}$. The values of $\tilde{g}$ are obtained from the measurement of the vertical frequency $\omega_z$ (see Supplemental Material).

\begin{figure}
\includegraphics[width=\columnwidth]{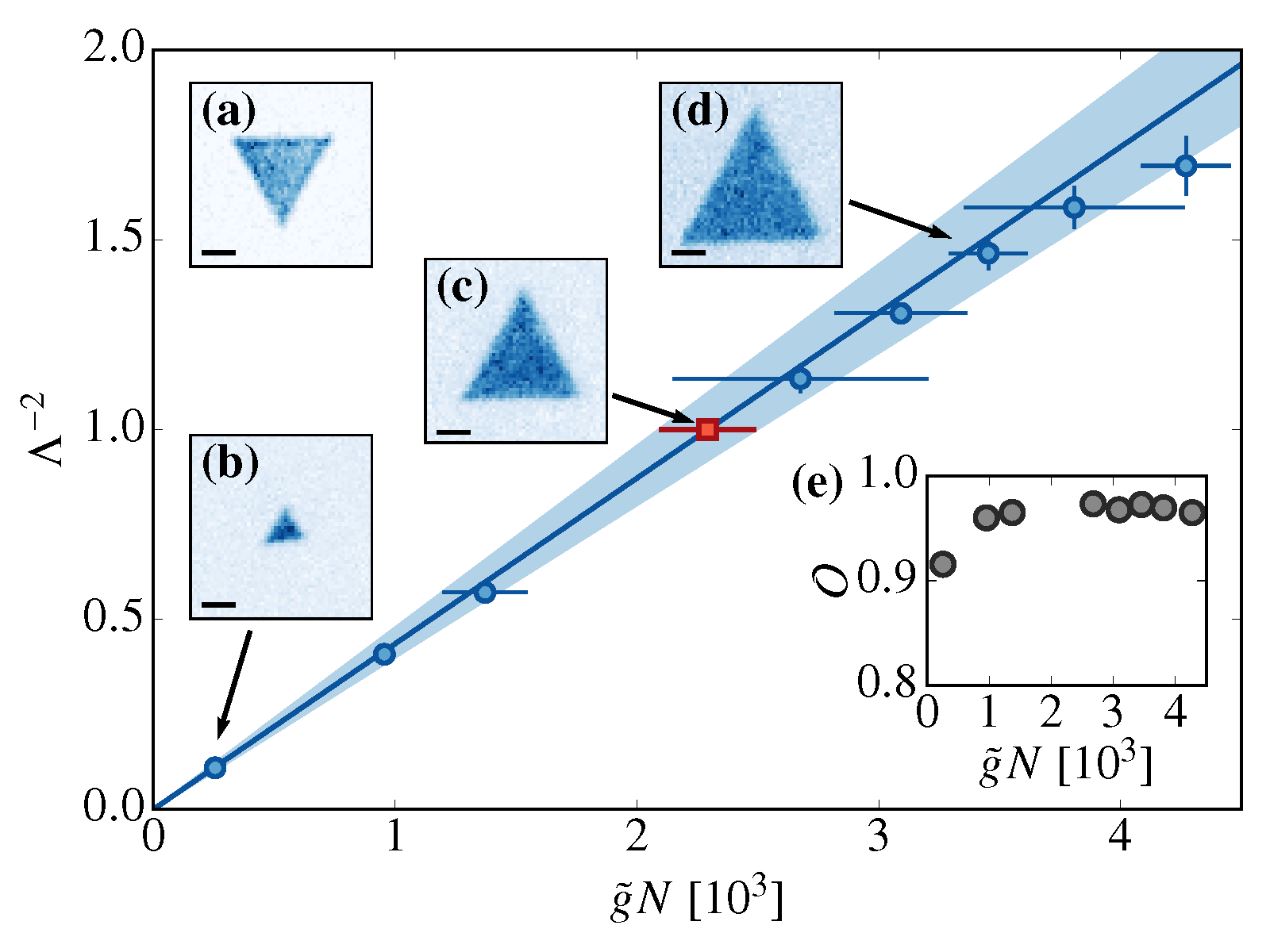}
\caption{Scaling factor at $\omega t = \pi/2$ for different values of $\tilde{g}N$. (a): Initial density distribution of the cloud. (b)-(d): Density distributions of the cloud after an evolution during $t = \pi/(2\omega)$ in the harmonic trap for different values of $\tilde{g}N$. For (a)-(d), the horizontal black lines represent $10\,\si{\micro\meter}$. Main graph: Best scaling factor $\Lambda^{-2}$ as a function of $\tilde{g}N$. The red square corresponds to the reference image and its ordinate is fixed to 1. The solid line represents the prediction (\ref{eq:scaling_gN}). The shaded area represents its uncertainty, due to the one in the atom number of the reference point. The vertical error bar represent the precision at two standard deviations of the fit that determines $\Lambda^{-2}$. (e): Value of the overlap between the density distributions and the reference point. The error bars due to the fit are smaller than the black points.}
\label{Fig3}
\end{figure}

We analyze the series of images using the same general method as in Sec.\,\ref{sec:same_Nat}. We select arbitrarily one image as a reference point (here, the one corresponding to $\tilde g N\approx2000$, shown as a red square on Fig. \ref{Fig3}). Then, we calculate the best overlap between this reference point and all other images obtained for different $\tilde g N$'s, and extract an optimal scaling parameter $\Lambda$. The results of this analysis are displayed on Fig.\,\ref{Fig3}. The inset shows that the overlap is close to 1 for all values of $\tilde g N$, indicating that the clouds all have the same shape, as expected from (\ref{eq:scaling_gN}). On the main graph of Fig.\,\ref{Fig3}, we show the variations of $\Lambda^{-2}$ with $\tilde g N$. The scaling law (\ref{eq:lambda_t_bis}) predicts that $\Lambda^{-2} = \mu^2\propto \tilde{g}N$, which is indicated by the solid line passing by the origin and the reference point. Here again, this prediction is in excellent agreement with the data. Note however that the result for the largest $\tilde g N$ is slightly lower than the theoretical prediction, which we attribute to the fact that, for larger powers in the vertical confining laser beam, the local defects of the potential that it creates start to play a significant role.

Interestingly, the shape for $t'=\pi/(2\omega)$, {i.e.}, $\ta=\infty$ for an evolution without any trap, is close to a uniformly filled triangle but inverted compared to the initial one (see insets of Fig.\,\ref{Fig3}). The emergence of such a simple form after time-of-flight is reminiscent of the simple diamond-like shape obtained for the 3D expansion of a uniform gas initially confined in a cylindrical box \cite{Gotlibovych:2014}. Note that we also observe such a diamond-like shape at $t=\pi/(2\omega)$ starting from a square box, albeit with a non-uniform density (see Supplemental Material).


\subsection{Connecting evolutions with a fixed trap frequency, different sizes and different $\tilde{g} N$\label{subsec:diff_size_Nat}}

\begin{figure}
\includegraphics[width=\columnwidth]{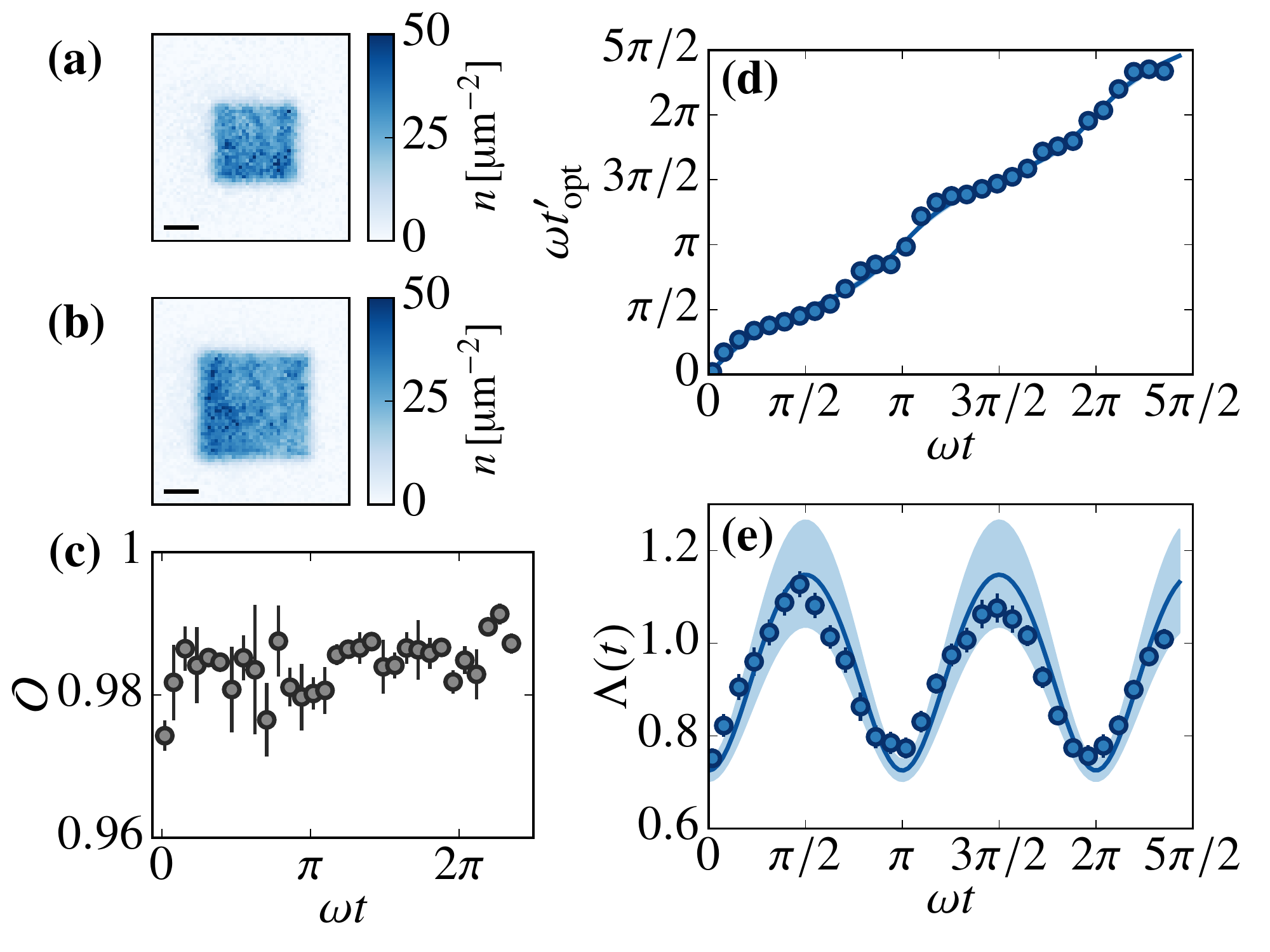}
\caption{Mapping between two clouds with the same shape, different sizes and different atom numbers. (a), (b): Initial density distribution of the two clouds. The horizontal black lines represent $10\,\si{\micro\meter}$. (c) Best overlap between each image of the first series of images and the images of the second one. (d): Optimal time $\tb_\mathrm{opt}$ of the second evolution as a function of the time $ \ta$ of the first evolution. (e): Optimal scaling factor $\Lambda(t)$ between the first and second evolutions. On (d) and (e), the solid lines are the predictions (\ref{eq:relation_t_tau_bis}) and (\ref{eq:lambda_t_bis}) where the values of the parameters $\pL$ and $\mu$ are measured independently. The uncertainty of these values are represented as a shaded area. On (d), this area is too narrow to be discernable. On (c)-(e), the error bars indicate the confidence intervals within two standard deviations of the fit that we use to reconstruct the scaling laws. They are too small to be seen on (d).}
\label{Fig4}
\end{figure}

Finally we compare the evolution of two clouds with homothetic shapes and $\pL,\mu \neq 1, \zeta=1$, which means clouds with different initial sizes, different atom numbers and evolving in the same harmonic trap. We perform an experiment where the initial shape is a square with a uniform density. The first cloud has a side length $L_1 = 27.0(5)\ \si{\micro\meter}$, contains $N_1 = 3.7(3)\times 10^4$ atoms, and its initial density distribution is shown on Fig.\,\ref{Fig4}(a). The second one has a side length $L_2 = 36.8(5)\ \si{\micro\meter}$ and contains $N_2 = 5.4(3)\times 10^4$ atoms (Fig.\,\ref{Fig4}(b)). The ratio $\xi/L$ is around $0.01$ for these two clouds. We let them evolve in the same harmonic potential described above and with the same interaction parameter $\tilde g$, and take pictures after different evolution times. We expect that the two evolutions $n_1(\ra,\ta)$ of the first cloud and $n_2(\rb,\tb)$ of the second cloud are linked via Eqs.\,(\ref{eq:transform_lambda3a}, \ref{eq:lambda_t_bis}, \ref{eq:relation_t_tau_bis}), with parameters $\pL=L_2/L_1 = 1.36(4)$ and $\mu = \sqrt{N_2/N_1}=1.21(8)$. We analyze the two series of images with the same procedure as in Sec.\,\ref{sec:same_Nat} and determine the scaling laws that link the two evolutions one to the other. The best overlaps between the images of the first and  second series are shown in Fig.\,\ref{Fig4}(c). They are all above 0.97, indicating that the two evolutions are indeed similar. The relation between the time $\tb$ of the second frame and the corresponding time $\ta$ of the first frame is shown on Fig.\,\ref{Fig4}(d), and the best scaling factor $\Lambda(\ta)$ is shown on Fig.\,\ref{Fig4}(e). The solid lines show the theoretical predictions (\ref{eq:relation_t_tau_bis}) and (\ref{eq:lambda_t_bis}), which are in very good agreement with the experimental data. 

With the three experiments described in Sec.\,\ref{sec:same_Nat} and Sec.\,\ref{sec:vary_gN}, we have tested the scaling laws (\ref{eq:transform_lambda3a})-(\ref{eq:relation_t_tau_bis}) independently for the three parameters $\pL$, $\mu$ and $\zeta$, demonstrating that, in the quantum hydrodynamic regime, the evolution of a cloud initially at rest depends only on its initial shape, up to scaling laws on space, time and atom density.


\section{Two-dimensional breathers}
\label{sec:breathers}

In Sec.\,\ref{sec:periodic_Ep}, we have seen that due to the SO(2,1) symmetry, the evolution of the potential energy $E_{\rm p}$ is periodic with period $T/2\equiv\pi/\omega$ for an arbitrary initial state $\psi(\bs r,0)$ [see Eq.\,(\ref{eq:Epot_vs_t})]. Of course, the existence of this periodicity does not put a strong constraint on the evolution of $\psi(\bs r,t)$ itself. Due to the non-linear character of the Gross-Pitaevskii equation, the evolution of $\psi$ is not expected to be periodic, as illustrated on Fig.\,\ref{fig:periodic_evol_Ep}(b) for a square initial shape. When looking experimentally or numerically at various initial shapes like uniformly-filled squares, pentagons or hexagons, we indeed observe that even though $E_{\rm p}(j T/2) = E_{\rm p}(0)$ for integer values of $j$, the shapes $n(\bs r)=N|\psi(\bs r)|^2$ at those times are notably different from the initial ones. We found two exceptions to this statement, which are the cases of an initial equilateral triangle and a disk. This section is devoted to the study of these very particular states that we call "breathers".  

\begin{figure*}[t]
\includegraphics[width=2\columnwidth]{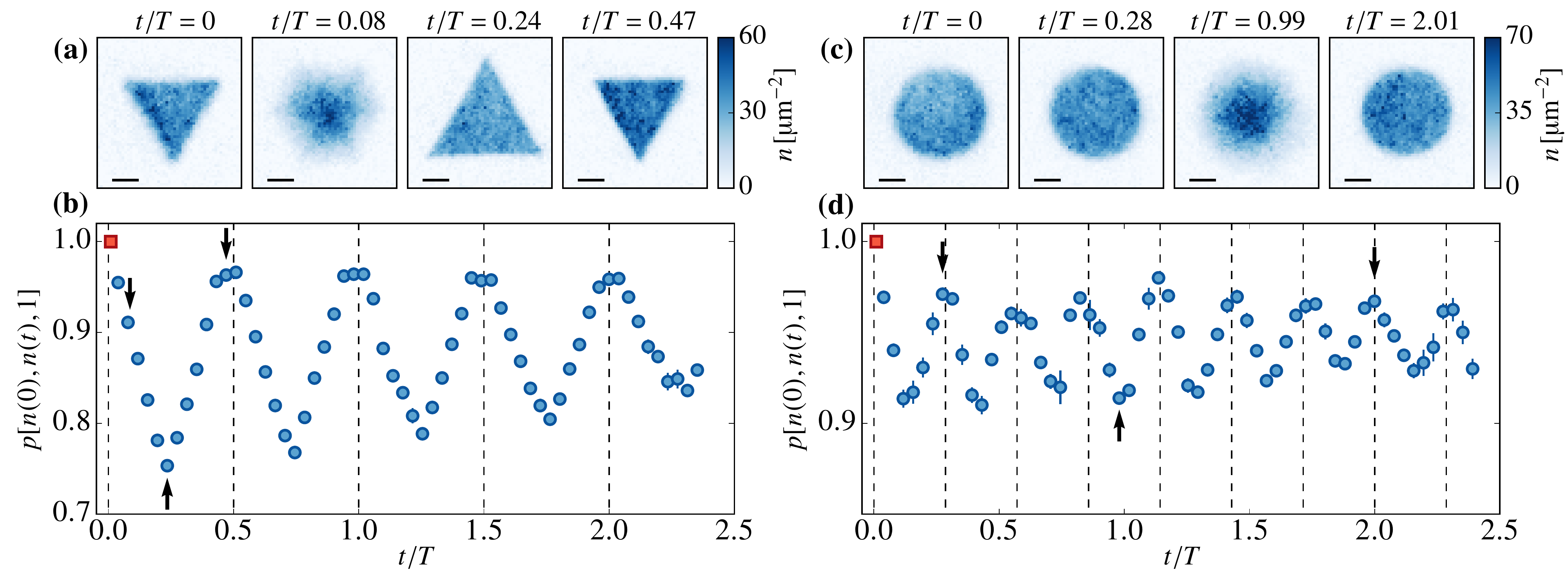}
\caption{(a): Density distributions of an initially triangular-shaped cloud at $t/T = 0$, $t/T = 0.08$, $t/T \approx 1/4$ and $t/T \approx 1/2$. The first and last distributions are close one to another. (b): Scalar product between the initial density distribution of a triangular-shaped cloud (red square) and the density distributions during its evolution in the harmonic trap. The first point is fixed at 1. The dashed lines indicate where $t/T$ is a multiple of $1/2$. The shape seems to be periodic of period $T/2$. (c): Density distributions of an initially disk-shaped cloud at $t/T = 0$, $t/T \approx 2/7$, $t/T \approx 1$ and  $t/T \approx 2$. The first two and the last distributions are close one to the others. (d): Scalar product between the initial density distribution of a disk-shaped cloud (red square) and the density distributions during its evolution in the harmonic trap. The first point is fixed at 1. The dashed lines indicate where $t/T$ is a multiple of $2/7$. The shape seems to be periodic of period $2/7$. On (a) and (c), the horizontal black lines represent $10\,\si{\micro\meter}$. On (b) and (d), the black arrows indicate the point corresponding to density distributions shown on (a) and (c) respectively. The error bars represent the statistical error of the measurement.}
\label{Fig5}
\end{figure*}

In the present context of a fluid described by the Gross--Pitaevskii equation, we define a breather as a wave function $\psi(\bs r,t)$ that undergoes a periodic evolution in an isotropic harmonic trap of frequency $\omega$ (for a generalisation to different settings, see e.g.\,\cite{Bishop:1980,dauxois2006physics}). According to this definition, the simplest example of a breather is a steady-state $\psi_{\rm ss}(\bs r)$ of the Gross--Pitaevskii equation, e.g. the ground state. Other breathers are obtained by superposing  $\psi_{\rm ss}$ with one eigenmode of the Bogoliubov--de Gennes equations, resulting from the linearization of the Gross-Pitaeveskii equation around $\psi_{\rm ss}$. In principle (with the exception of the breathing mode \cite{Pitaevskii:1997a}), the population of this mode should be vanishingly small to avoid damping via non-linear mixing. Extending this scheme to the superposition of several modes in order to generate more complex types of breathers seems difficult. Indeed the eigenmode frequencies are in general non-commensurable with each other, therefore the periodicity of the motion cannot occur as soon as several modes are simultaneously excited \footnote{For the ground state of a harmonically confined 2D gas in the Thomas--Fermi limit, the mode frequencies are 
$\omega\left(2n^2+2nm+2n+m\right)^{1/2}$ with $n,m$ positive or null integers \cite{Stringari:1998,Ho:1999}.}. Note that for a negative interaction coefficient $\tilde g$ in 1D, a bright soliton forms a stable steady-state of the Gross--Pitaevskii equation (even for $\omega\to 0$) and thus also matches our definition. In that particular 1D case, a richer configuration exhibiting explicitly the required time-periodicity is the Kuznetsov--Ma breather, which is obtained by superposing a bright soliton and a constant background (see e.g.\,\cite{Zhao:2018} and refs.\,in).

Here, we are interested in 2D breathers that go well beyond a single mode excitation and we start our study  with the uniform triangular shape. In this case, for experiments performed with a gas in the Thomas--Fermi regime, we find that the evolution of the shape is periodic with period $T/2$ within the precision of the measurement. As an illustration, we show in Fig.\,\ref{Fig5}(a) four images taken between $t=0$ and $T/2$. The scalar product $(n(0)|n(t))$ between the initial distribution and the one measured at times $T/2$, $T$, $3T/2$ and $2T$, shown in Fig.\,\ref{Fig5}(b), is indeed very close to 1. We could reproduce the same result for various initial atom numbers. 

\begin{figure*}[t]
\includegraphics[width=2\columnwidth]{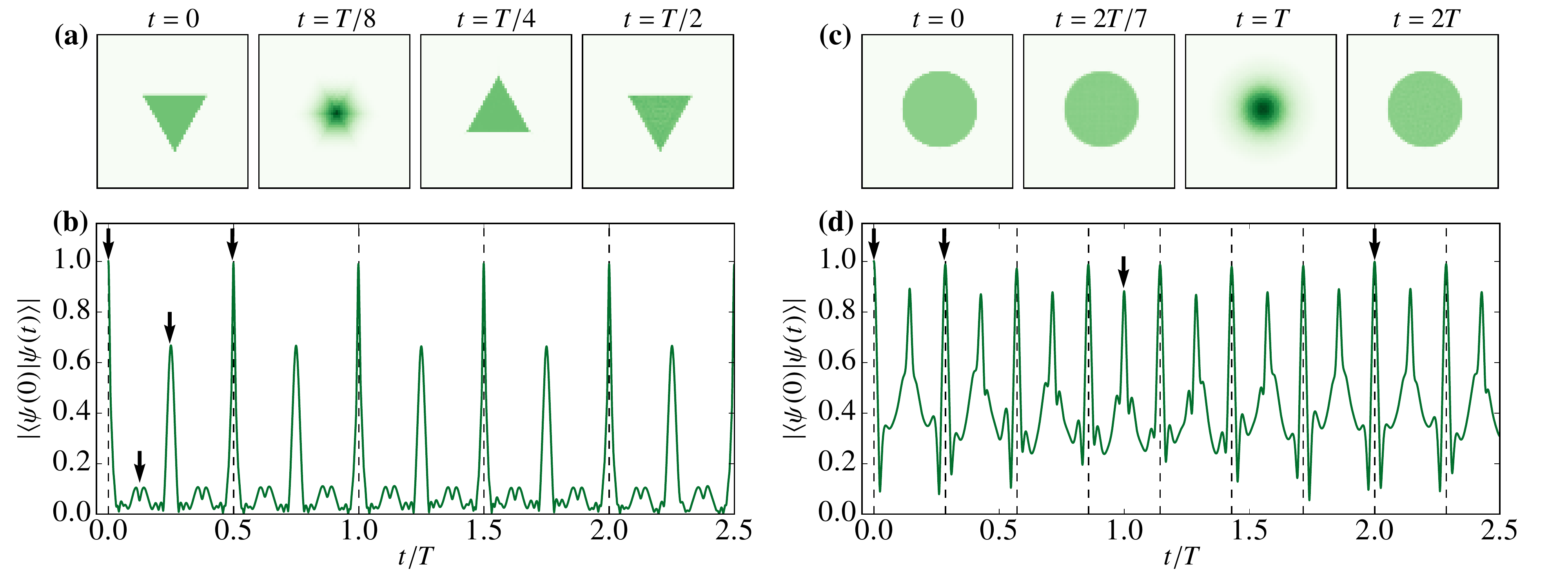}
\caption{(a): Calculated density distributions at times $t/T=0,\,1/8,\,1/4,\,1/2$ and (b): calculated time evolution of $|\langle \psi(0)|\psi(t)\rangle|$, starting from the ground state in a triangular box. The numerical integration of the Gross--Pitaevskii equation is performed on a $512\times 512$ grid. The triangle is centered on the grid, with a side length equal to half the grid size. We chose $\tilde g N=25600$ corresponding to an initial healing length $\xi\approx \ell$, where $\ell$ is the grid step. (c): Calculated density distributions at times $t/T=0,\,2/7,\,1,\,2$ and (d): calculated time evolution of $|\langle \psi(0)|\psi(t)\rangle|$, starting from the ground state in a disk-shaped box. The numerical integration of the Gross--Pitaevskii equation is performed on a $512\times 512$ grid. The disk is centered on the grid, with a diameter equal to half the size of the grid. We chose $\tilde g N=12800$ leading to an initial healing length $\xi\approx 2\ell$, where $\ell$ is the grid step. On (b) and (d), the black arrows indicate the times corresponding to the snapshots presented on (a) and (c).}
\label{triangle_512}
\end{figure*}

We did not find an analytical proof of this remarkable result, but we could confirm it numerically by simulating the evolution of a wave function $\psi(\bs r,t)$ with the Gross--Pitaevskii equation \footnote{In spite of the fact that our parameters are well in the Thomas--Fermi regime, we perform the numerical analysis using the Gross--Pitaevskii equation (\ref{eq:GPeq}) and not the quantum hydrodynamic equations (\ref{eq:continuity2},\ref{eq:GPE2_approx}). The reason is that the discontinuity of the density that appears in the latter case on the edge of the sample may lead to numerical singularities in the subsequent dynamics.}. 
We show in Fig.\,\ref{triangle_512}(a) a few snapshots of the calculated density distribution, and in Fig.\,\ref{triangle_512}(b) the evolution of the modulus of the (usual) scalar product $|\langle \psi(0)|\psi(t)\rangle|$ between the wave functions at times $0$ and $t$. The calculation was performed on a square grid of size $N_s\times N_s$ with $N_s=512$. The initial wave function is the ground state of a triangular box with the side length $N_s/2$ centered on the grid, obtained by imaginary time evolution for $\tilde g N=25600$. Note that by contrast to the "scalar product between images" introduced above, the quantity $|\langle \psi(0)|\psi(t)\rangle|$ is also sensitive to phase gradients of the wave functions. Its evolution shows clear revivals approaching unity for $t$ close to multiples of $T/2$.

\begin{figure}[t]
\includegraphics[width=\columnwidth]{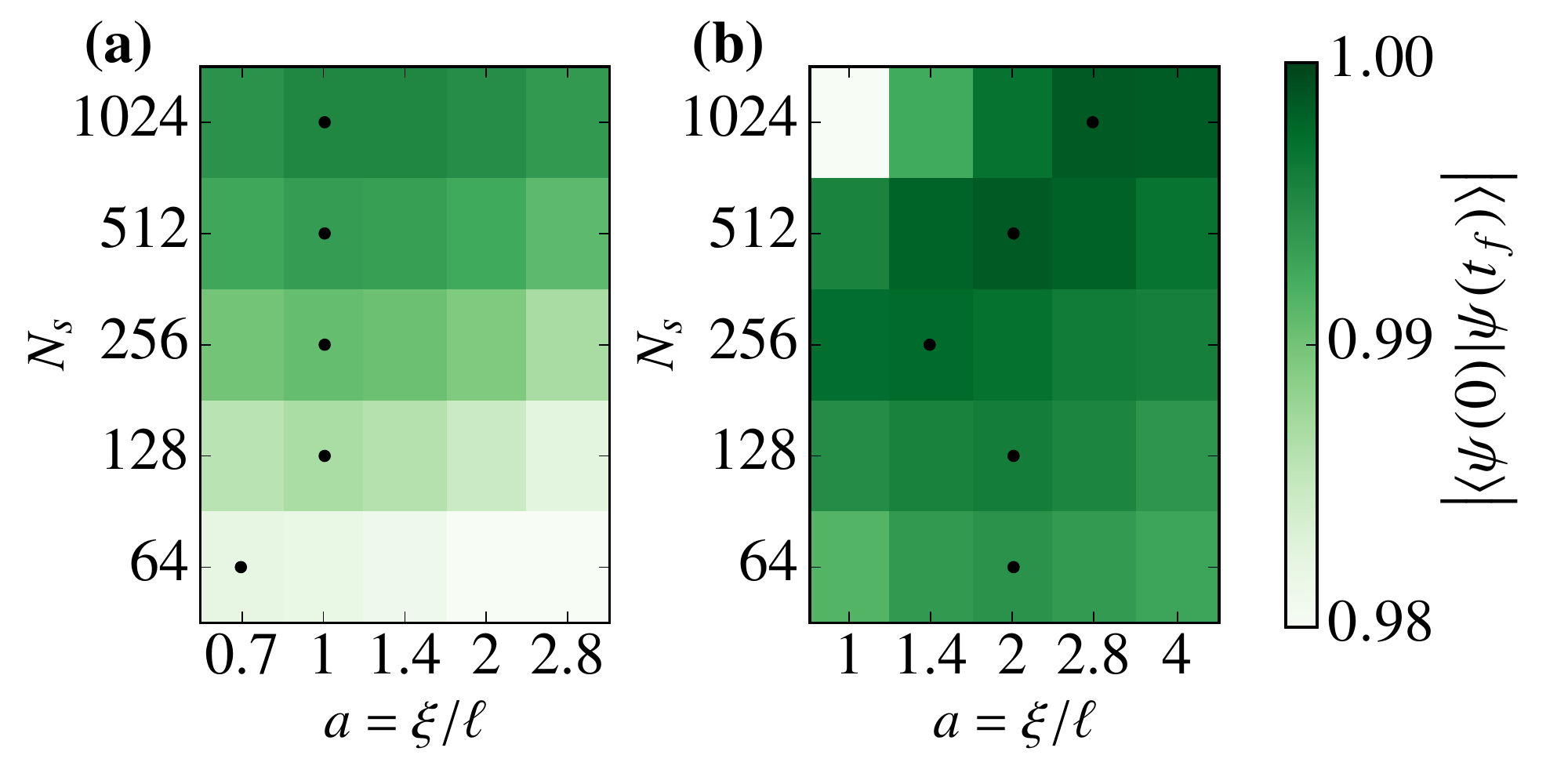}
\caption{Finite-size scaling for the numerical simulations. (a) Scalar product $|\langle \psi(0)|\psi(T/2)\rangle|$ for an initial triangular shape. The size of the grid $N_s$ and the sampling of the healing length $a\equiv\xi/\ell$ are varied. The highest value is 0.9953, obtained for $N_s = 1024, a = 1$. (b) Scalar product $|\langle \psi(0)|\psi(2T)\rangle|$ for an initial disk shape. The highest value is 0.9986, obtained for $N_s = 1024, a = 2.8$. On both figures, the black dots indicate the highest value of the scalar product for each line. }
\label{Fig6}
\end{figure}

We show in Fig.\,\ref{Fig6}(a) the finite-size scaling analysis of the value of the first maximum of this scalar product occurring at $t_{\rm max}\approx T/2$, for increasing grid sizes $N_s=64,\ldots,1024$.  The product $\tilde gN$ is adjusted such that the healing length $\xi = [N\hbar^2/(2mE_{\rm i})]^{1/2}= a\ell$, where $\ell$ is the grid spacing and $a^2 = 0.5,1,2,4,8$. The condition $a\ll N_s$ ensures that $\xi$ is much smaller than the size of the triangle (Thomas--Fermi regime), while having $a \gtrsim 1$ provides an accurate sampling of the edges of the cloud. The overlap between $|\psi(0)\rangle$ and $|\psi(t_{\rm max})\rangle$ increases with the grid size, and reaches $0.995$ for the largest grid.

In the simulation, the trapping frequency $\omega$ is adjusted such that $|\Delta E| \ll E_{\rm tot}$ in Eq.\,(\ref{eq:Epot_vs_t}); the cloud then keeps an approximately constant area over time, which is favourable for the numerics. Note that this choice does not restrict the generality of the result, since the scaling laws seen in Sec.\,\ref{sec:same_Nat} allow one to connect the evolution of a given $\psi(\bs r,t=0)$ in traps with different frequencies. In particular, if the evolution starting from $\psi(\bs r,0)$ in a trap of frequency $\omega_1$ is periodic with period $\pi/\omega_1$, the evolution in another trap with frequency $\omega_2$ will be periodic with period $\pi/\omega_2$ (see Eq.\,(\ref{eq:relation_t_tau})).

Two simulations with the same ratio $a/N_s \propto \xi/L$, where $L = \ell N_s/2$ is the size of the initial cloud, describe the same physical system with a better accuracy as $a$ and $N_s$ are increased. For the results in Fig.\,\ref{Fig6}(a), increasing the number of pixels $N_s$ for a fixed $a/N_s$ makes the scalar product closer to $1$. If this result could be extended as such to arbitrary large values of $N_s$, this would demonstrate that the ground state of a triangular box evolves periodically in a harmonic potential. However, a closer look at the results of this finite-size scaling analysis   seems to indicate that $a$ should  either be kept constant or increased at a slower rate than $N_s$ to have the scalar product approaching 1 in an optimal way. Of course this conjecture deduced from our numerical analysis needs to be further explored with analytical tools, which is out of the scope of the present paper.

The requirement for the Thomas-Fermi regime ($\xi/L \ll 1$) is necessary for obtaining a periodic evolution of the shape with period $T/2$. Indeed, in the ideal gas case ($\tilde g=0$), the evolution over $T/2$ corresponds to an inversion of the initial shape with respect to the origin, {i.e.}, a triangle pointing upwards for the case of interest here (Fig.\,\ref{Fig5}(a)). One may then wonder about the existence of a  periodicity $T$ for the triangular shape, irrespective of the product $\tilde gN$. Indeed this periodicity holds in both limiting cases $\tilde g=0$ (ideal gas) and $\tilde gN$ large (Thomas--Fermi regime). However numerical simulations  show unambiguously that the evolution is not periodic in the intermediate case.

We have also run the same simulations for other simple regular polygons (square, pentagons, hexagon). We did not observe a similar revival of the initial wave function over the time period $(0,5T)$ (see Supplemental Material for details).

Finally we turn to the case of a disk-shaped initial cloud (Fig.\,\ref{Fig5}(c)). The experiment was performed with a cloud prepared such that $|\Delta E|\ll E_{\rm tot}$ in Eq.\,(\ref{eq:Epot_vs_t}), so that the potential energy is approximately constant over time. In this particular case, the experimental result shown in Fig.\,\ref{Fig5}(d) seems to indicate a periodicity $\approx 2T/7$ for the evolution of the overlap between $n(\bs r,0)$ and $n(\bs r,t)$. To illustrate this, Fig.\,\ref{Fig5}(c) displays four density distribution at times between $0$ and $2T$. Let us assume that this periodicity $2T/7$ is exact when $\Delta E=0$. For a disk-shaped initial distribution with any value of $\Delta E$, the evolution cannot be $2T/7$-periodic. Indeed the potential energy of the cloud is only $T/2$-periodic, which is not a submultiple of $2T/7$. However, all the disk-shaped clouds should have a $2T$ periodicity, which is the least common multiple of $T/2$ and $2T/7$. As we show now, this $2T$ periodicity is well supported by a numerical analysis.


We show in figure \ref{triangle_512}(c) snapshots of the the calculated density distribution, and in figure \ref{triangle_512}(d) the time evolution of the overlap $|\langle \psi(0)|\psi(t)\rangle|$, starting from the ground state in a disk-shaped box potential centered on a $512\times 512$ grid. The disk diameter is chosen equal to half the grid size and the simulation is run for $\tilde g N=12800$. This simulation shows that the overlap between $\psi(\bs r,0)$ and $\psi(\bs r,t)$ indeed recovers values close to 1 at times close to multiples of $2T/7$, as observed experimentally. 

A closer inspection of figure \ref{triangle_512}(d) indicates that the time evolution of the overlap is in good approximation periodic with period $2T$, with a symmetry around $t=T$ as well as around $t=2T$. If the evolution is effectively periodic with period $2T$, the symmetry around these points is expected. Indeed the wave function is chosen real for $t=0$, and will thus be real also at $2T$ (up to a global phase). Therefore the evolution must be symmetric around those points thanks to time-reversal symmetry. On the other hand, this symmetry does not show up around the other local maxima $j 2T/7$ ($j=1,\ldots,6$), indicating that one does not expect a full overlap with the initial state for those points.  

In order to investigate further the revival around $2T$, we have run a finite-size scaling analysis for the same grid sizes as for the triangles, and for $a^2 = 1,2,4,8,16$ (Fig.\,\ref{Fig6}(b)).  We find that the numerical results are compatible with a full recovery of the initial wave function at time $2T$, with a scalar product between the wave functions at times $0$ and $2T$ attaining a maximum of $0.9986$ for the largest grid size $N_s=1024$ and $a^2=8$. In this case, the optimal value of $a$ for a given $N_s$ (marked with a dot in Fig.\,\ref{Fig6}) increases with $N_s$; note that the optimal ratio $a/N_s\propto \xi/L$ decreases when $N_s$ increases, which guarantees that the cloud remains in the Thomas-Fermi regime. 

To conclude this section, we emphasize that the phenomenon described here is notably different from the existence of a breathing mode at frequency $2\omega$ \cite{Kagan:1996b,Pitaevskii:1997a}, mentioned in the introduction and explored in Sec.\,\ref{sec:periodic_Ep}. Here, we observe a periodic motion of the whole cloud, not just of the second moment $\langle r^2\rangle$ of the position. We also note that the observed phenomenon is a genuine non-linear effect, which cannot be captured by a linearization of the motion of the cloud around an equilibrium position. Indeed the state of the gas at an intermediate time may dramatically differ from the state at initial time or after a full period, both in terms of size and shape. A proper analysis of these breathers may require a multimode approach, with the observed phenomenon resulting from a mode synchronization effect via non-linear couplings. 


\section{Summary and outlook}

In this paper, we investigated experimentally some important consequences of the dynamical symmetries of the two-dimensional Gross-Pitaevkii equation, describing the evolution of a weakly interacting Bose gas in a harmonic potential. Firstly, we showed that the SO(2,1) symmetry leads to a periodic evolution of the potential energy and to scaling laws between the evolution of clouds with the same atom number and the same interaction parameter.
Secondly, we showed that in the quantum hydrodynamic regime, more symmetries allow one to describe  the evolution of the gas by a single universal function irrespective of its size, atom number, trap frequency, and interaction parameter $\tilde{g}$. This universal evolution depends only on the initial shape and velocity field of the cloud.
Thirdly, we identified two geometrical box-like potentials, equilateral triangle and disk, which lead to a periodic motion of the wave function when one starts with a gas uniformly filling these shapes and releases it in a harmonic potential of frequency $\omega$. The period of these breathers are $\pi/\omega$ and $4\pi/\omega$ for the triangles and the disks, respectively. This result was confirmed by a numerical simulation for a cloud initially in the Thomas-Fermi regime of the box-like potential, giving an overlap respectively larger than 0.995 and 0.998 between the initial state and the state after one period.

The existence of these breathers raises several interesting questions. First it is not immediate that their existence is a direct consequence of the dynamical symmetries of the system. If this is the case, such breathers could appear also for other systems exhibiting the SO(2,1) symmetry, like a three-dimensional unitary Fermi gas or a cloud of particles with a $1/r^2$ interaction potential. Remarkably the later case can be approached using classical (Newton) equations of motion; we have performed a preliminary numerical analysis with up to $10^5$ particles, which indicates that an initial triangular (resp. disk) shape with uniform filling also leads to an approximate periodic evolution in a harmonic potential with same period $T/2$ (resp. $2T$) as the solution of the Gross--Pitaevskii equation. We also note that in the 1D case, the spectrum of the Hamiltonian of a gas of particles interacting with a repulsive $1/r^2$ potential is composed of evenly spaced energy levels, ensuring a periodic evolution of the system for any initial state \cite{Calogero:1971,Sutherland:1972}. 

The allowed shapes for such breathers is also an intriguing question. In our exploration (both experimental and numerical), we only found this behavior for triangles and disks but one cannot exclude that complex geometrical figures can show a similar phenomenon. Another issue is related with thermal effects. For all studies reported here, we operated with a gas deeply in the degenerate regime, which is well approximated by the zero-temperature Gross--Pitaevskii formalism. A natural extension of our work is therefore to study to which extent the present findings will subsist in the presence of a significant non-superfluid component. For our experimental setup, this will require a significant increase in the vertical trapping frequency, so that the vertical degree of freedom remains frozen for the thermal component of the gas.

Finally we recall that the SO(2,1) symmetry is only an approximation for the description of a two-dimensional Bose gas. It is valid when the gas can be modelled by a classical field analysis, hence  for a small interaction parameter $\tilde g\ll 1$. For stronger interactions, one has to turn to a quantum treatment of the fluid. This breaks the scale invariance and the SO(2,1) symmetry that exist at the classical field level, providing an example of a "quantum anomaly" \cite{Holstein:1993,Cabo:1998,Olshanii:2010}. For example, the frequency of the breathing mode of a gas in a harmonic potential then differs from its classical value $2\omega$. It remains to be understood if a similar "quantum anomaly" shows up for the breathers described in this work.

\begin{acknowledgments}
This work is supported by DIM NanoK, ERC (Synergy UQUAM), QuantERA ERA-NET (NAQUAS project) and the ANR-18-CE30-0010 grant. We thank Yvan Castin, Cheng Chin, Ignacio Cirac, Lei Feng, J\"org Schmiedmayer and Steven Simon for stimulating discussions.
\end{acknowledgments}


\appendix

\section{Symmetry groups of the Schr\"odinger and 2D Gross--Pitaevskii equations \label{app:SO(2,1)}}

For completeness we summarize in this appendix the main properties of the transformations that leave invariant the Schr\"odinger equation (i) for a free particle and (ii) for a particle confined in a harmonic potential. The ensemble of these transformations forms a group called the \emph{maximal kinematical invariance group}, which is parametrized in the 2D case by 8 real numbers. In what follows, we are interested only in the subgroup that is relevant for scale/conformal invariance. For example in the case of a free particle, 5 parameters are related to space translations, changes of Galilean frame and rotations, which do not play a role in our study. We are then left with 3 parameters, corresponding to time translations, dilations and special conformal transformations. These transformations also leave  the 2D Gross-Pitaevskii equation invariant. In the following we identify their generators and show that they obey the SO(2,1) commutation algebra. We will follow closely the approach of \cite{Niederer:1972,Niederer:1973}, which was developed for the Schr\"odinger equation describing the motion of a single particle, but also applies with little modifications to the case of the non-linear Gross-Pitaevskii equation. In this appendix we set $\hbar=1$ to simplify the notations.

\subsection{Free particles}
\label{subsec:appendix_free_particle}
Although we are interested ultimately in the case where the particles evolve in a harmonic potential, we start by a brief summary of the free particle case, for which the algebra is slightly simpler, while involving transformations of a similar type. 
In \cite{Niederer:1972}, it was shown that in addition to space translations, rotations, and Galilean transformations, the three following transformations leave invariant the free-particle Schr\"odinger equation:
\begin{itemize}
 \item
The translations in time 
\begin{equation}
\bs r \to\bs r,\qquad t \to t+\beta,
\end{equation}
since the Hamiltonian has no explicit time dependence.

 \item
The dilations
\begin{equation}
\bs r \to\bs r/\lambda,\qquad t \to t/\lambda^2,
\label{eq:dilation}
\end{equation}
already introduced in Eq. (1) of the main text.

\item 
The so-called "expansions":
\begin{equation}
\bs r \to \frac{\bs r}{\gamma t+1},\qquad \bs t \to \frac{t}{\gamma t+1},
\label{}
\end{equation}
which correspond to a special conformal transformation for the time.

\end{itemize} 
The combination of these transformations forms a 3-parameter group with the most general transformation written as:
\begin{eqnarray}
\bs r \to g(\bs r,t)\equiv\frac{\bs r}{\gamma t + \delta}
\label{eq:general1}\\
\bs t \to h(t)\equiv\frac{\alpha t + \beta}{\gamma t + \delta},
\label{eq:general2}
\end{eqnarray}
with the constraint $\alpha\delta - \beta\gamma = 1$. The dilation (\ref{eq:dilation}) is obtained by setting $\beta=\gamma=0$, $\delta=\lambda$ and $\alpha\delta=1$.

Let us consider a function $\psi_1(\bs r,t)$ which is a solution of the Gross-Pitaevskii equation in free space:
\begin{equation}
{\cal P}_0[\psi_1;\bs r,t]=0
\end{equation}
with
\begin{equation}
{\cal P}_0[\psi;\bs r,t]\equiv \I \frac{\partial \psi}{\partial t}+\frac{1}{2m}\bs \nabla^2_{\bs r} \psi -\frac{\tilde gN}{m}|\psi|^2\psi.
\end{equation}
 
Starting from $\psi_1(\bs r,t)$, we define the function $\psi_2(\bs r',t')$ as 
\begin{equation}
\psi_2(\bs r',t')=f(\bs r,t)\, \psi_1(\bs r,t)
\label{eq:psi2_psi1}
\end{equation}
with $\bs r',t'$ set as 
\begin{equation}
\bs r'=g(\bs r,t), \ t'=h(t)
\label{eq:rp_r_tp_t}
\end{equation}
and
\begin{equation}
f(\bs r,t)=\left(\gamma t+\delta\right)\exp\left(-\I \frac{m\gamma r^2/2}{\gamma t + \delta}  \right).
\end{equation}
With a tedious, but straightforward calculation, one can check that $\psi_2 (\bs r',t')$ is also a solution of the Gross--Pitaevskii equation:
\begin{equation}
{\cal P}_0[\psi_2;\bs r',t']=0
\end{equation} 
for any value of the parameters $\alpha, \beta, \gamma, \delta$ with the constraint $\alpha\delta - \beta\gamma = 1$. The group of transformations (\ref{eq:general1}, \ref{eq:general2}) thus allows one to generate an infinite number of solutions of the Gross-Pitaevskii equation. We could pursue this analysis by determining the generators associated with the action of these transformations on the wave functions $\psi(\bs r,t)$, but we postpone it to the case of a harmonically confined system which is more relevant for our physical system. The two studies are anyway very similar and the symmetry groups of the two systems have the same Lie algebra \cite{Niederer:1972,Niederer:1973}.

\subsection{Particles in a harmonic trap}
\label{subsec:appendix_harmonic}
In the presence of an isotropic harmonic potential of frequency $\omega$, the general transformations on position and time leaving invariant the Schr\"odinger equation are also defined by a set of four numbers $(\alpha,\beta,\gamma,\delta)$ with the constraint $\alpha\delta-\beta\gamma=1$ \cite{Niederer:1973}. Setting
\begin{equation}
\eta=\tan(\omega t),\ \eta'=\tan(\omega t'),
\end{equation}
the change in position is
\begin{equation}
\bs r \to \bs r'=g(\bs r,t)\equiv \frac{\bs r}{\lambda(t)}\\
\label{eq:transform_r_appendix}
\end{equation} 
with
\begin{eqnarray}
\lambda(t)&=&\left[\left( \alpha \sin(\omega t)+\beta\cos(\omega t)\right)^2 \right.
\nonumber \\
&+& \left. \left( \gamma \sin(\omega t)+\delta\cos(\omega t) \right)^2\right]^{1/2},
\label{eq:lambda_t_general_appendix}
\end{eqnarray}
while the transformation on time $t \to t'=h(t)$ reads
\begin{equation}
\eta'=\frac{\alpha \eta +\beta}{\gamma \eta +\delta}.
\label{eq:transform_t_appendix}
\end{equation}
Note that the time translations belong to this set of transformations, as expected for a time-independent problem. They are obtained by taking $\alpha=\delta=\cos(\omega t_0)$ and $\beta=-\gamma=\sin(\omega t_0)$.

We start with a solution $\psi_1$ of the Gross-Pitaevskii equation in the trap:
\begin{equation}
{\cal P}_\omega[\psi_1;\bs r,t]=0
\end{equation}
with
\begin{equation}
{\cal P}_\omega[\psi;\bs r,t]={\cal P}_0[\psi;\bs r,t]-\frac{1}{2}m\omega^2r^2 \psi.
\end{equation}
Using this group of transformations, we can generate another function $\psi_2(\bs r',t')$ satisfying 
\begin{equation}
{\cal P}_\omega[\psi_2;\bs r',t']=0,
\end{equation}
following the definitions (\ref{eq:psi2_psi1}-\ref{eq:rp_r_tp_t}) with now
\begin{equation}
f(\bs r,t)=\lambda(t)\;\exp\left(-\I\frac{m \dot \lambda r^2}{2\lambda}\right).
\end{equation}
The fact that $\psi_2$ is a solution of the Gross-Pitaevskii equation was proven for the non-interacting case in \cite{Niederer:1973}, and one can check that the contribution of the interaction term proportional to $|\psi|^2\psi$ cancels in the 2D case thanks to the scaling $f\propto \lambda$.

In the main text, we use a specific version of the transformation $(\bs r,t)\to (\bs r',t')$ that (i) maps the time $t=0$ onto the time $t'=0$, and (ii) is such that $\dot \lambda(0)=0$ since we want to relate a real solution $\psi_1$ onto another real solution $\psi_2$ ($\psi_1$ and $\psi_2$ are both ground state wave functions in a box-like potential). These two conditions, in association with $\alpha\delta-\beta \gamma=1$, impose $\beta=\gamma=0$ and $\delta=1/\alpha$, hence:
\begin{equation}
\lambda(t)=\left[\alpha^2 \sin^2(\omega t)+\frac{1}{\alpha^2}\cos^2(\omega t) \right]^{1/2}
\label{eq:lambda_t_simplified}
\end{equation}
and
\begin{equation}
\tan(\omega t')=\alpha^2\,\tan(\omega t).
\label{eq:tan_t_simplified}
\end{equation}

Finally we note that the simple dilation transformation $\bs r' =\bs r/\sqrt{\zeta}$, $t'= t/\zeta$ allows one to relate a solution of the Gross-Pitaevskii equation $\psi_1(\bs r,t)$ in a trap with frequency $\omega_1$ to a solution 
\begin{equation}
\psi_2(\bs r',t')=\sqrt{\zeta}\,\psi_1(\bs r,t)
\end{equation}
in a trap with frequency $\omega_2=\zeta\omega_1$:
\begin{equation}
{\cal P}_{\omega_1}[\psi_1;\bs r,t]=0\ \Rightarrow\ {\cal P}_{\omega_2}[\psi_2;\bs r',t']=0.
\end{equation}
We can thus combine this dilation with the transformation (\ref{eq:lambda_t_simplified}-\ref{eq:tan_t_simplified}) in order to obtain the transformation that links two (initially real) solutions $\psi_{1}(\bs r,t)$ and $\psi_2(\bs r',t')$ of the Gross--Pitaevskii equation for a given $\tilde gN$, obtained in harmonic traps with frequencies $\omega_{1,2}$ and starting with homothetic initial conditions with characteristic lengths $L_{1,2}$. This transformation reads:
\begin{equation}
\bs r'=\frac{\bs r}{\lambda(t)},\qquad \tan(\omega_2 t')=\zeta\alpha^2 \tan(\omega_1 t)
\end{equation}  
with
\begin{equation}
\lambda(t)=\left[\alpha^2\zeta^2 \sin^2(\omega_1 t)+\frac{1}{\alpha^2}\cos^2(\omega_2 t) \right]^{1/2}
\end{equation}
and $\alpha=L_2/L_1$, $\zeta=\omega_2/\omega_1$. This corresponds to the scaling (\ref{eq:lambda_t}) used in the main text.

\subsection{Generators and SO(2,1) symmetry}

We now look for the infinitesimal generators of the transformation $\psi_1\to \psi_2$ in the presence of a harmonic potential (Sec.\,\ref{subsec:appendix_harmonic}), and show that they fulfill the commutation algebra characteristic of the SO(2,1) group. We focus here on the transformation (\ref{eq:transform_r_appendix}-\ref{eq:transform_t_appendix}) which relates solutions of the Gross--Pitaevskii equation for the same non-linear coefficient $\tilde g N$ and the same trap frequency $\omega$. 

We first note that the set of four numbers $(\alpha,\beta,\gamma,\delta)$ with the constraint $\alpha\delta-\beta\gamma=1$ actually forms a set of three independent parameters, as for the free-particle case (Sec.\,\ref{subsec:appendix_free_particle}). To this set of numbers we can associate a matrix 
\begin{equation}
M=\begin{pmatrix} \alpha & \beta \\ \gamma &\delta \end{pmatrix}
\end{equation}
of the group SL(2,{R}). In order to simplify our discussion, we consider the following three subgroups of SL(2,{R}), each parametrized by a single parameter $s_j$, $j=1,2,3$:
\begin{equation}
\begin{pmatrix} \E^{s_1/2} & 0 \\ 0 &\E^{-s_1/2} \end{pmatrix},\quad 
\begin{pmatrix} \cosh(s_2/2) & \sinh(s_2/2) \\ \sinh(s_2/2) &\cosh(s_2/2) \end{pmatrix},\ 
\end{equation}
and 
\begin{equation}
\begin{pmatrix} \cos(s_3/2) & -\sin(s_3/2) \\ \sin(s_3/2) &\cos(s_3/2) \end{pmatrix}.
\end{equation} 
We will obtain three independent generators by considering a small displacement from the unit matrix for each subgroup ($|s_j|\ll 1$). In all three cases, we will write the passage from $\psi_1$ to $\psi_2$ as:
\begin{equation}
\psi_2(\bs r,t)\approx \left( \hat 1-\I s_j  \hat L_j(t)\right)  \psi_1(\bs r,t), 
\label{eq:infinitesimal_psi}
\end{equation}
where we introduce the time-dependent generator $\hat L_j(t)$. The goal is to determine explicitly these operators and their commutation relation, in order to check that they satisfy the SO(2,1) algebra.
 
\paragraph{Generator associated to $s_1$.} We have in this case
\begin{equation}
M \approx \hat 1+ \frac{s_1}{2}\hat \sigma_z,
\label{eq:gene1}
\end{equation}
where the $\hat \sigma_j$, $j=x,y,z$, are the Pauli matrices.
We first get $\lambda(t)=1-\frac{s_1}{2}\cos(2\omega t)$ so that 
\begin{equation}
f(\bs r,t) = 1 - \frac{s_1}{2}\cos(2\omega t) - \I s_1 \frac{m\omega r^2}{2}\sin(2\omega t),
\end{equation}
and the infinitesimal changes in $\bs r,t$ are
\begin{equation}
g(\bs r,t)\approx \bs r \left(1+\frac{s_1}{2}\cos(2\omega t)\right), \quad h(t)=t+\frac{s_1}{2\omega}\sin(2\omega t).
\label{eq:g_h_case_1}
\end{equation}
This allows one to determine the passage from $\psi_1$ to $\psi_2$ as in (\ref{eq:infinitesimal_psi}) with
\begin{eqnarray}
\hat L_1(t)&=&-\frac{\I}{2} \cos(2\omega t) \left(1+\bs r\cdot \bs \nabla\right) 
\nonumber \\
&+&\frac{1}{2\omega} \sin(2\omega t) \left(
m\omega^2 r^2-\I\partial_t \right).
\end{eqnarray}

\paragraph{Generator associated to $s_2$.} We find
\begin{equation}
M \approx \hat 1+ \frac{s_2}{2}\hat \sigma_x.
\label{eq:gene2}
\end{equation}
In this case $\lambda(t)=1+\frac{s_2}{2}\sin(2\omega t)$, and
\begin{equation}
f(\bs r,t) = 1 + \frac{s_2}{2}\sin(2\omega t) - \I s_2 \frac{m\omega r^2}{2}\cos(2\omega t).
\end{equation}
It also provides the transformation of space and time coordinates:
\begin{equation}
g(\bs r,t)\approx \bs r \left(1-\frac{s_2}{2}\sin(2\omega t)\right), \quad h(t)=t+\frac{s_2}{2\omega}\cos(2\omega t).
\end{equation}
This corresponds to a transformation similar to the one considered above in Eq. (\ref{eq:g_h_case_1}), with the time translation $t\to t+\pi/(4\omega)$. 
The associated operator for the passage from $\psi_1$ to $\psi_2$ is thus
\begin{eqnarray}
\hat L_2(t)&=&\frac{1}{2\omega}\cos(2\omega t) \left(m\omega^2 r^2 
-\I\partial_t \right) \nonumber \\ 
&+& \frac{\I}{2} \sin(2\omega t) \left(1+\bs r\cdot \bs \nabla \right).
\end{eqnarray}

\paragraph{Generator associated to $s_3$.} Finally we have for the third case
\begin{equation}
M \approx \hat 1- \frac{s_3}{2}\I \hat \sigma_y.
\label{eq:gene3}
\end{equation}
We simply have $\lambda(t)=1$, $f(\bs r,t) = 1$, and this case corresponds to the time translations mentioned above, for which we have
\begin{equation}
g(\bs r,t)= \bs r , \quad h(t)=t-{s_3}/{2\omega}.
\end{equation}
 The operator $\hat L_3(t)$ is thus
\begin{equation}
\hat L_3(t)= \frac{\I}{2\omega}\partial_t.
\end{equation} 
 
From the expressions of the three generators $\hat L_j$ we easily find the commutations relations valid at any time
\begin{equation}
[\hat L_1,\hat L_2]=-\I  \hat L_3, \ \
[\hat L_2,\hat L_3]=\I  \hat L_1,\ \
[\hat L_3,\hat L_1]=\I  \hat L_2,
\end{equation} 
which are characteristic of the Lorentz group SO(2,1). As explained in \cite{Pitaevskii:1997a}, this set of commutation relations  allows one to construct in particular families of solutions with an undamped breathing motion.

%

\newpage
\null
\newpage
%
%
%
%
%

\renewcommand\thefigure{S\arabic{figure}}

\begin{center}
\LARGE
Supplemental material for:\\ 
Dynamical symmetry and breathers in a two-dimensional Bose gas
\end{center}

 



\subsection*{S1: Measurement of the vertical frequency \label{app:vert_freq}}

\begin{figure}[h]
\includegraphics[width=\columnwidth]{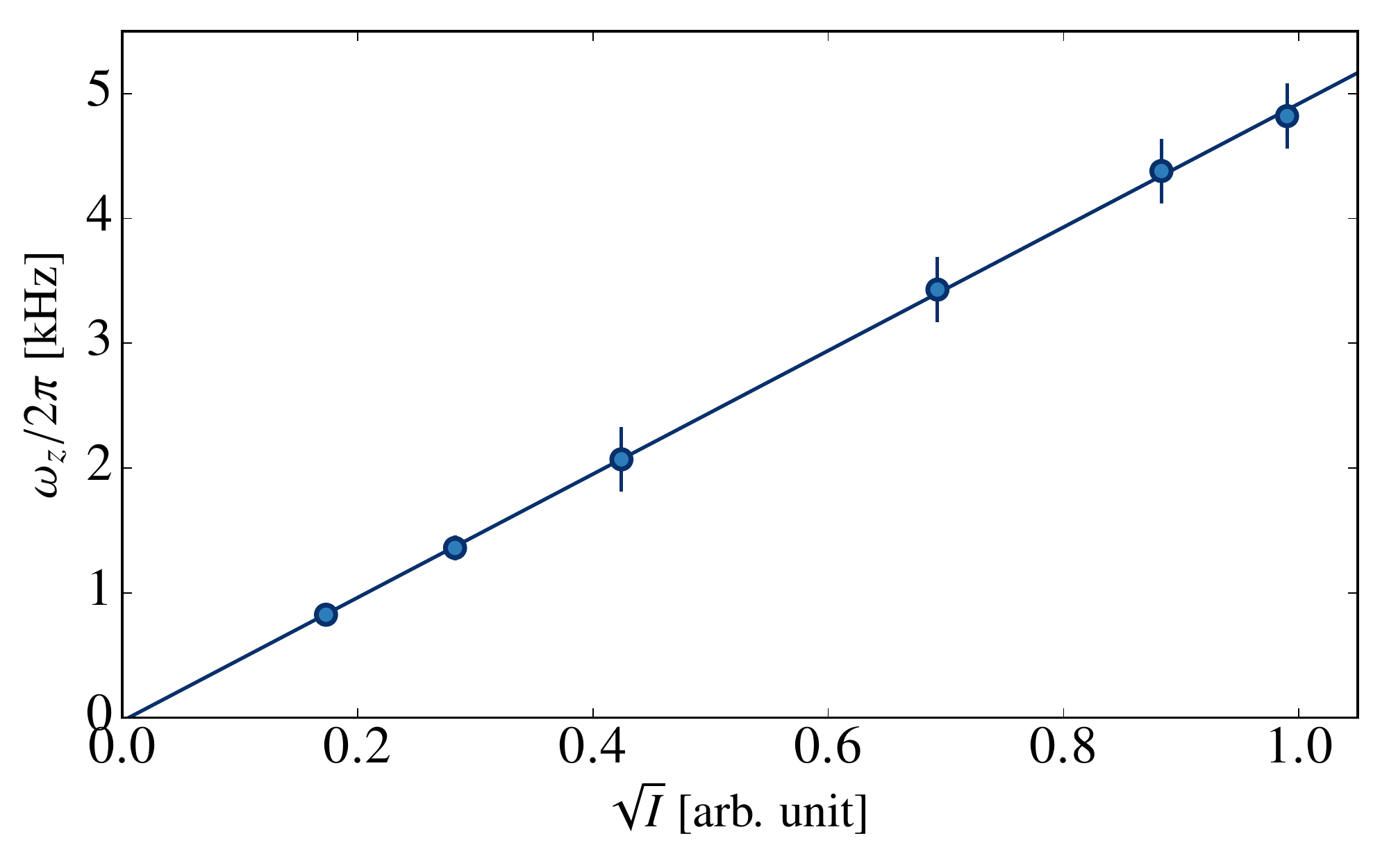}
\caption{Measurement of the vertical frequency $\omega_z$ for different values of the intensity of the laser creating the vertical confinement. All the experiments described above are realized in the range of intensities explored here and this curve enables us to determine the vertical frequency and the value of $\tilde{g}$ for all of these experiments.}
\label{fig:omega_z}
\end{figure}

The value of the interaction parameter $\tilde{g} = \sqrt{8\pi}a_s/\ell_z$ is determined by the measurement of the vertical frequency $\omega_z$, since $\ell_z = \sqrt{\hbar/(m\omega_z)}$. We measure this frequency for different values of the intensity of the laser producing the vertical confinement with the following procedure. We prepare a cloud with a high intensity of the laser, which means a high  vertical confinement ($\omega_z/(2\pi) \approx 7\,\si{\kilo\hertz}$), and in $1\,\si{\milli\second}$ we reduce this intensity to a final value $I$, giving a vertical kick to the cloud due to gravity and exciting a dipolar oscillation along the vertical axis. We let it evolve during a variable amount of time, then remove all confinements to let the cloud expand freely during $12\,\si{\milli\second}$. We measure its vertical position that varies sinusoidally with time. A fit of this vertical position provide us an estimate of $\omega_z$. On Fig. \ref{fig:omega_z} are shown the results of these measurements as a function of $\sqrt{I}$. The points lie on a line as expected, and a linear fit gives $\omega_z/(2\pi) = 4.95\,\si{\kilo\hertz}\times\sqrt{I} - 0.03\,\si{\kilo\hertz}$, where $\sqrt{I}$ is expressed in the arbitrary unit used on Fig. \ref{fig:omega_z}. We are then able to extrapolate the value of the vertical frequency, and therefore of $\tilde{g}$ for any intensity $I$ in the range explored on the figure.

\subsection*{S2: Details on the determination of the scaling laws\label{app:details_mapping}}

\begin{figure}[h]
\includegraphics[width=\columnwidth]{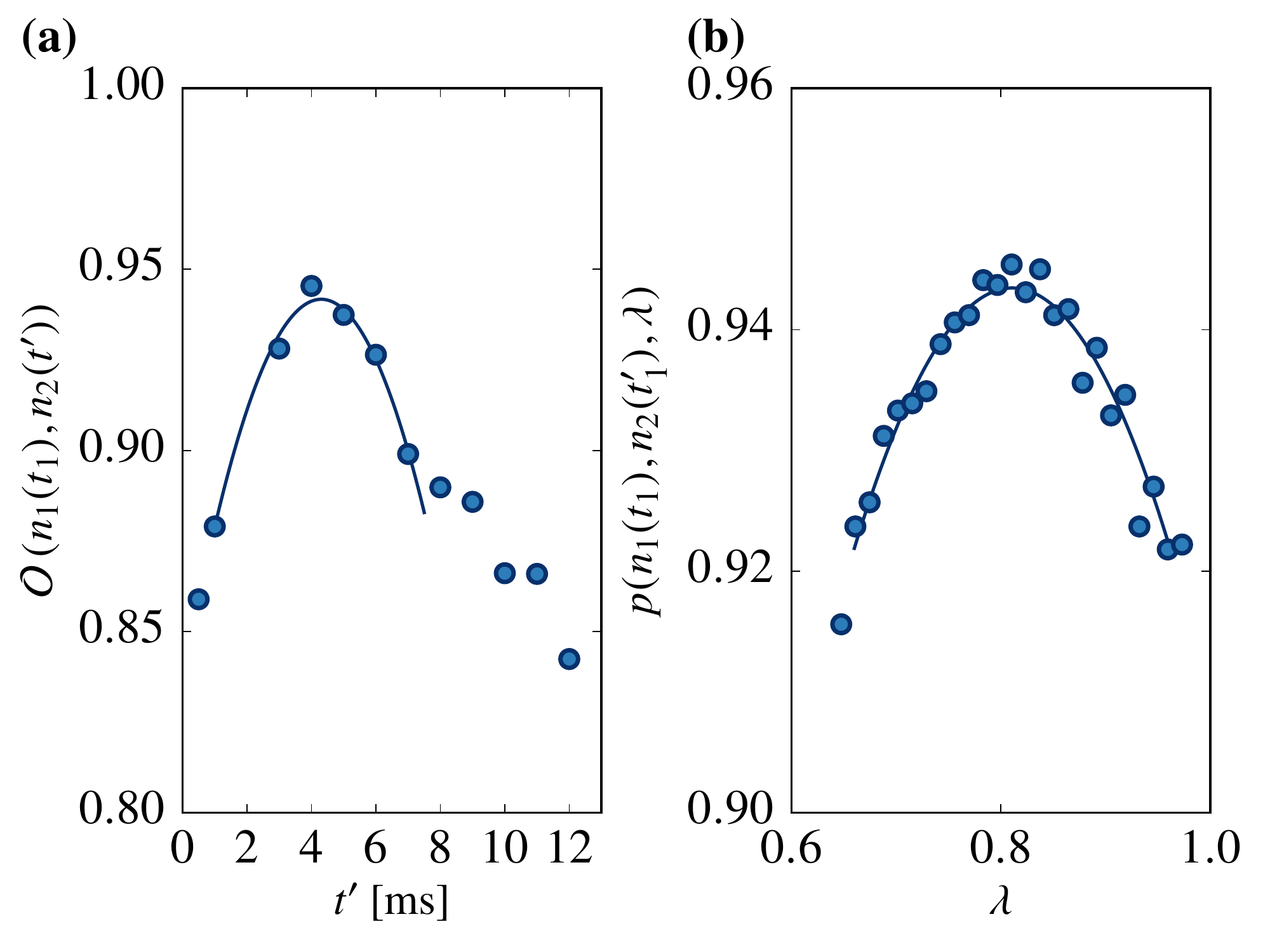}
\caption{Determination of the optimal time $t'_\mathrm{opt}(t_1)$ and of the optimal scaling factor $\Lambda(t_1)$ for the example of $t_1 = 5.9\,\si{\milli\second}$. (a) Overlap between $n_1(t_1)$ and $n_2(t')$ as a function of t'. The parabolic fit determines the time $t'_\mathrm{opt}(t_1)$ where the maximum is reached. (b) Scalar product between $n_1(t_1)$ and $n_2(t'_1)$ for various scaling factors $\lambda$. The parabolic fit determines the value of $\Lambda(t_1)$.}
\label{fig:overlap}
\end{figure}

We develop here with an example the procedure to reconstruct the scaling laws that we have used in sections III and IV\ C. We consider the density distribution $n_1(t_1)$ in the series presented on Fig. 2(a) at time $t_1 = 5.9\,\si{\milli\second}$, and we search for the time $t'_\mathrm{opt}(t_1)$ and the rescaling parameter $\Lambda(t_1)$ that relate this density distribution to the series presented on Fig. 2(b). We show on Fig. \ref{fig:overlap}(a) the overlap between the image $n_1(t_1)$ and all the images $n_2(t')$ of the second series. The optimal time is found to be $t_\mathrm{opt}(t_1) = 4.3(7)\,\si{\milli\second}$, indicated by the maximum of the parabolic fit. The time $t'_1$ that is the closest to $t'_\mathrm{opt}(t_1)$ for which we have an image is $t' = 4\,\si{\milli\second}$. On Fig. \ref{fig:overlap}(b) we show the variation with $\lambda$ of the scalar product $p\left(n_1(t_1),n_2(t'_1),\lambda\right)$. The optimal parameter is $\Lambda(t_1) = 0.81(5)$, indicated by the parabolic fit.

\subsection*{S3: Time evolution of density distributions in a harmonic trap (experiment)\label{app:time_evolution}}

\begin{figure*}[t]
\includegraphics[width=2\columnwidth]{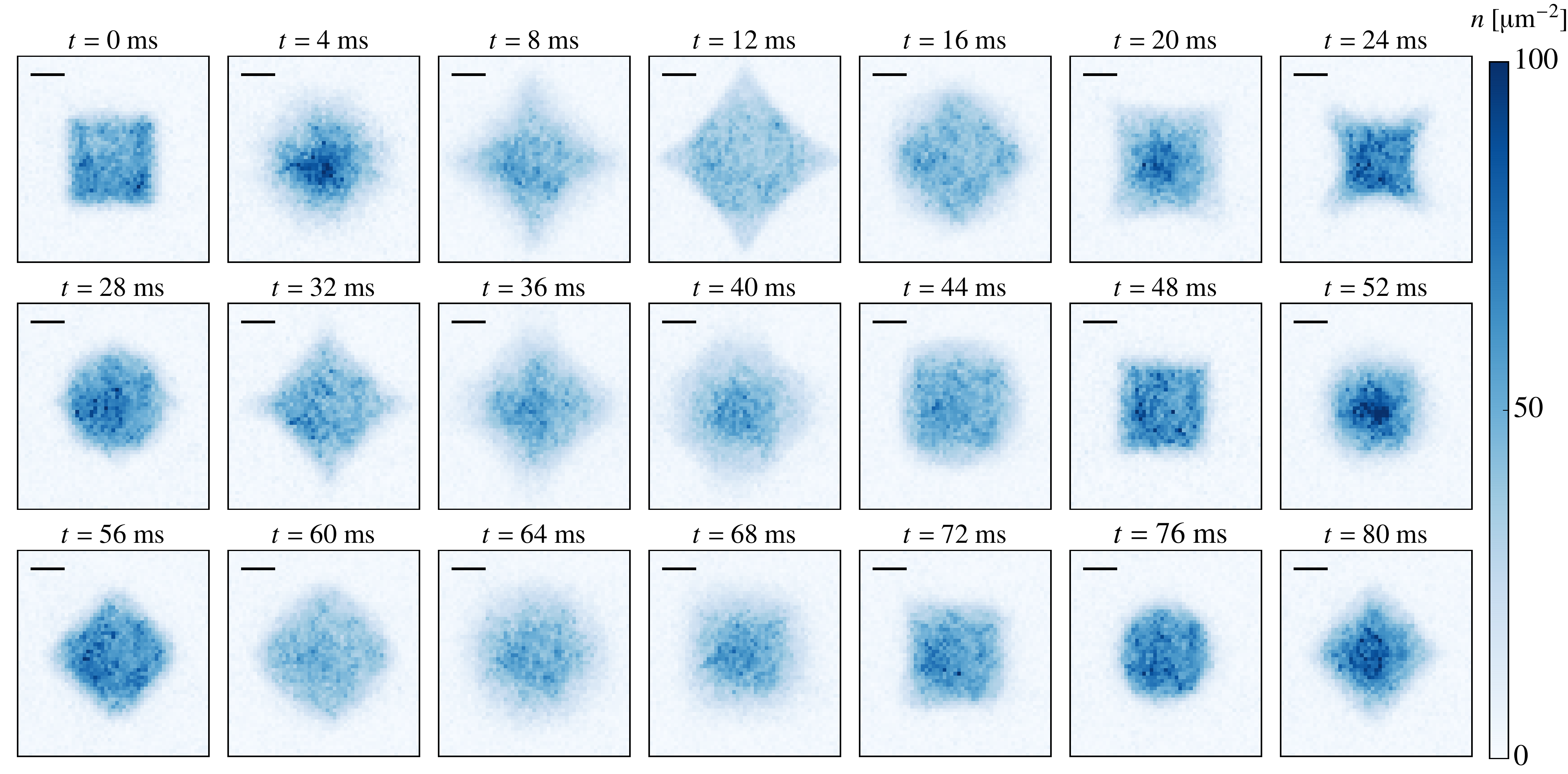}
\caption{Time evolution of the density distribution of the cloud presented on Fig. 1. The initial distribution has a shape of a square with side length $27.6(5)\,\si{\micro\meter}$, and the harmonic potential has a frequency of $\omega/(2\pi) = 19.3(1)$\,Hz. Each image is an average over 7 to 10 experimental realizations. The horizontal black lines represent $10\,\si{\micro\meter}$.}
\label{fig:time_evolution_detail}
\end{figure*}

The time evolution of the cloud whose potential energy is shown on Fig. 1 is displayed on Fig. \ref{fig:time_evolution_detail}. The imaging is performed by transfering a fraction of the atoms in the state $F=2,m_F=0$ and an absorption imaging is performed with a laser beam resonant with the $F=2 \to F'=3$ transition. The fraction of the transfered atoms is chosen so that the optical density of the imaged atoms doesn't exceed 1. The value of the atomic density is then obtained with
$n = \mathrm{OD}/(\varepsilon\sigma)$,
where $\mathrm{OD}$ is the measured optical density, $\varepsilon$ is the fraction of the transfered atoms and $\sigma$ is the scattering cross-section of the atoms.

On Fig. \ref{fig:time_evolution_detail}, the image at $\omega t = \pi/2$ corresponds to $t = 12\,\si{\milli\second}$, and we recover the observation that an initial square shape evolves as a diamond-shaped cloud, as discussed in the main text (Sec. IV B).

\subsection*{S4: Time evolution of density distributions in a harmonic trap (numerical)\label{app:time_evolution2}}

\begin{figure*}[th]
\includegraphics[width=2\columnwidth]{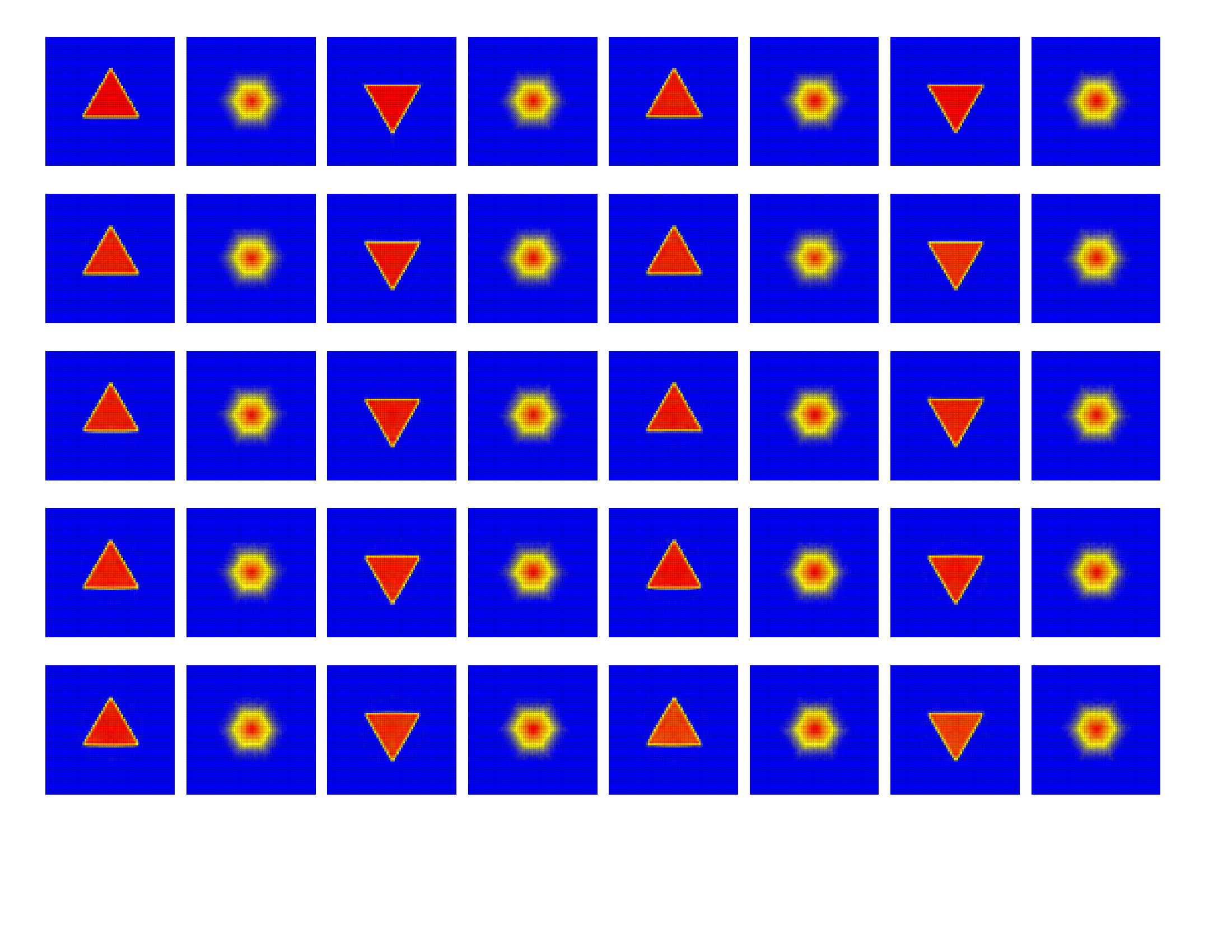}
\caption{Time evolution of the density $|\psi(\bs r,t)|^2$ calculated from the solution of the Gross--Pitaevskii equation for an initial triangular box potential (see text for details). Each image correspond to time $t=jT/8$, with $j=0,\ldots,39$. The evolution is approximately periodic with period $T/2$. }
\label{fig:evol_triangle}
\end{figure*}

\begin{figure*}[th]
\includegraphics[width=2\columnwidth]{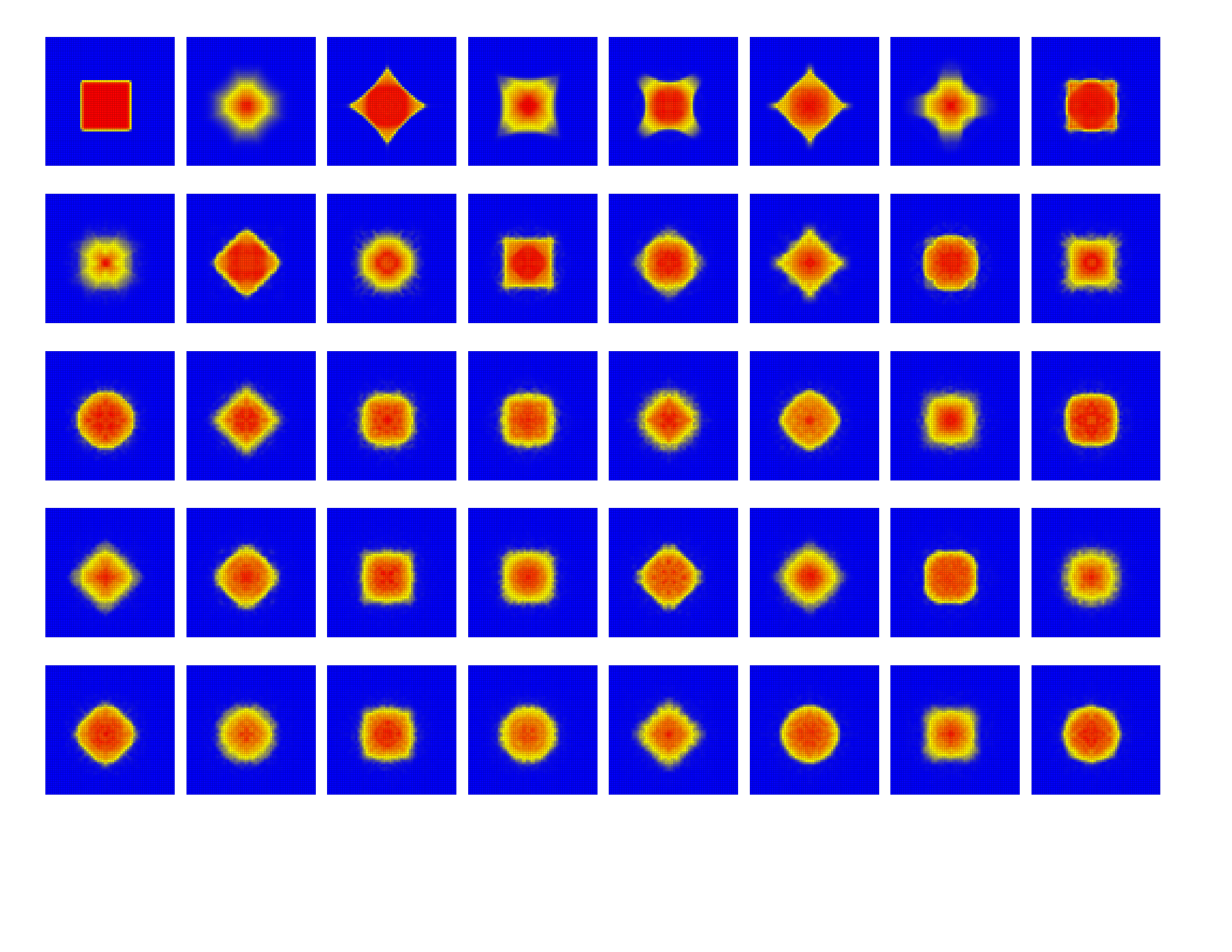}
\caption{Time evolution of the density $|\psi(\bs r,t)|^2$ calculated from the solution of the Gross--Pitaevskii equation for an initial square  box potential (see text for details). Each image correspond to time $t=jT/8$, with $j=0,\ldots,39$. No periodicity appears for this time interval. }
\label{fig:evol_square}
\end{figure*}

\begin{figure*}[p]
\includegraphics[width=2\columnwidth]{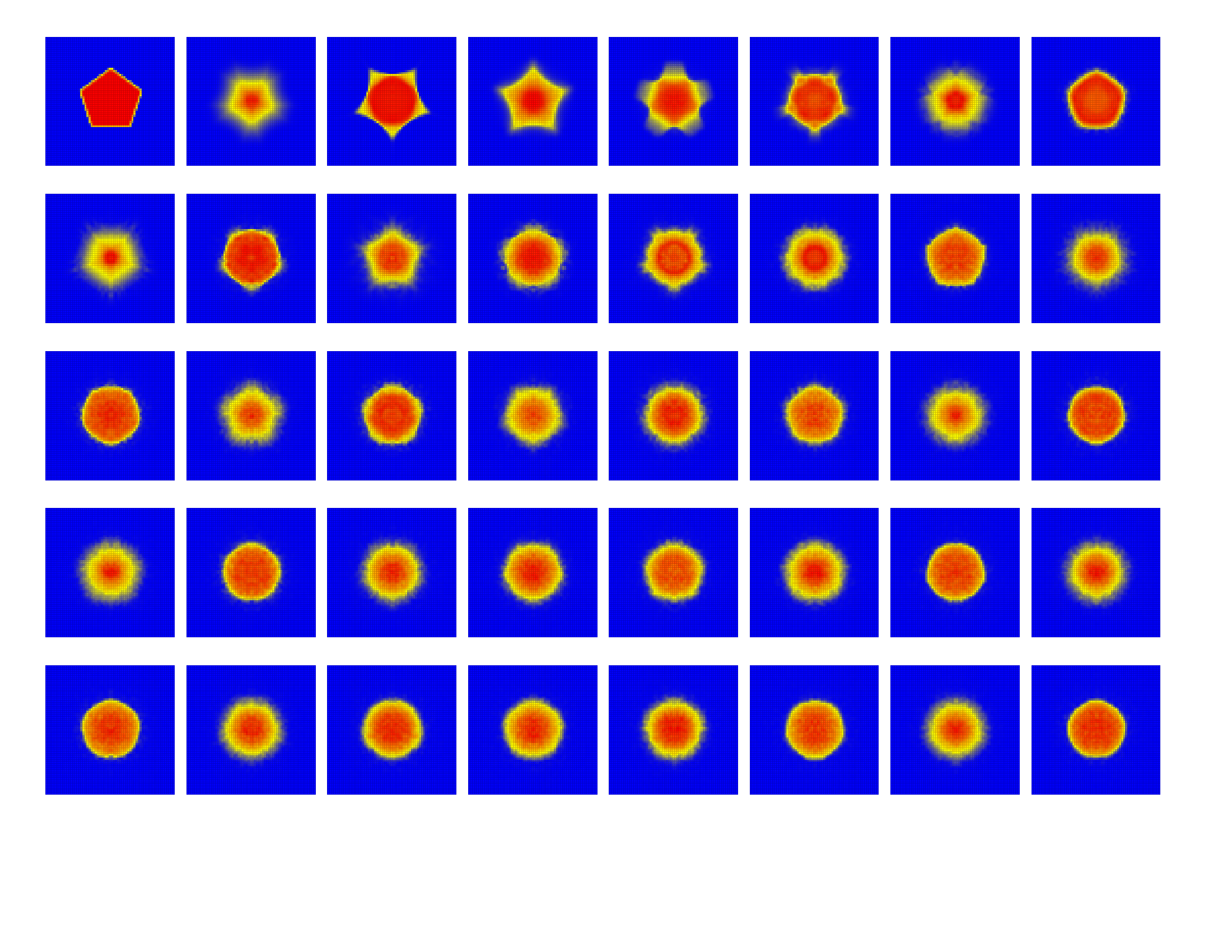}
\caption{Time evolution of the density $|\psi(\bs r,t)|^2$ calculated from the solution of the Gross--Pitaevskii equation for an initial pentagonal  box potential (see text for details). Each image correspond to time $t=jT/8$, with $j=0,\ldots,39$. No periodicity appears for this time interval. }
\label{fig:evol_pentagon}
\end{figure*}

\begin{figure*}[p]
\includegraphics[width=2\columnwidth]{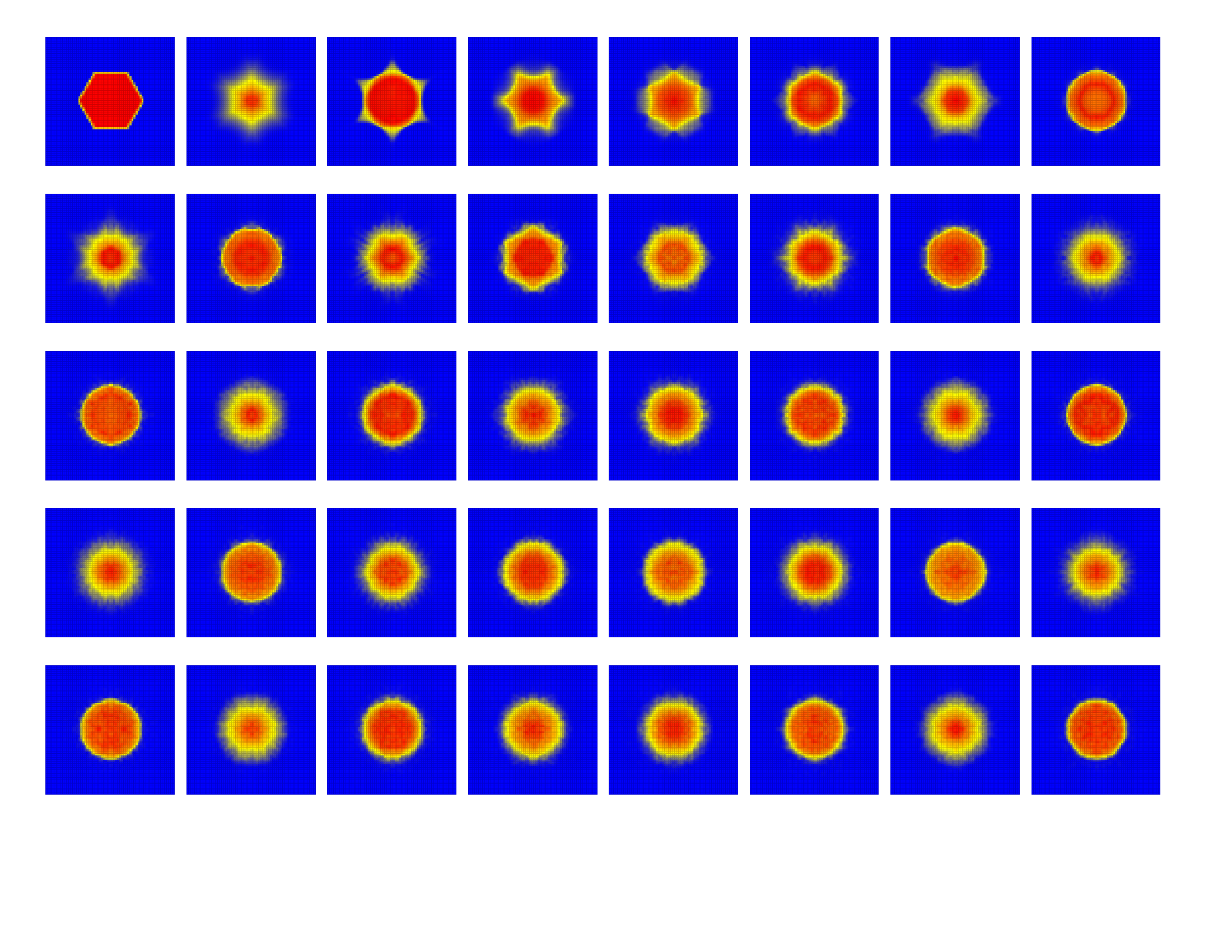}
\caption{Time evolution of the density $|\psi(\bs r,t)|^2$ calculated from the solution of the Gross--Pitaevskii equation for an initial hexagonal  box potential (see text for details). Each image correspond to time $t=jT/8$, with $j=0,\ldots,39$. No periodicity appears for this time interval. }
\label{fig:evol_hexagon}
\end{figure*}

\begin{figure*}[p]
\includegraphics[width=2\columnwidth]{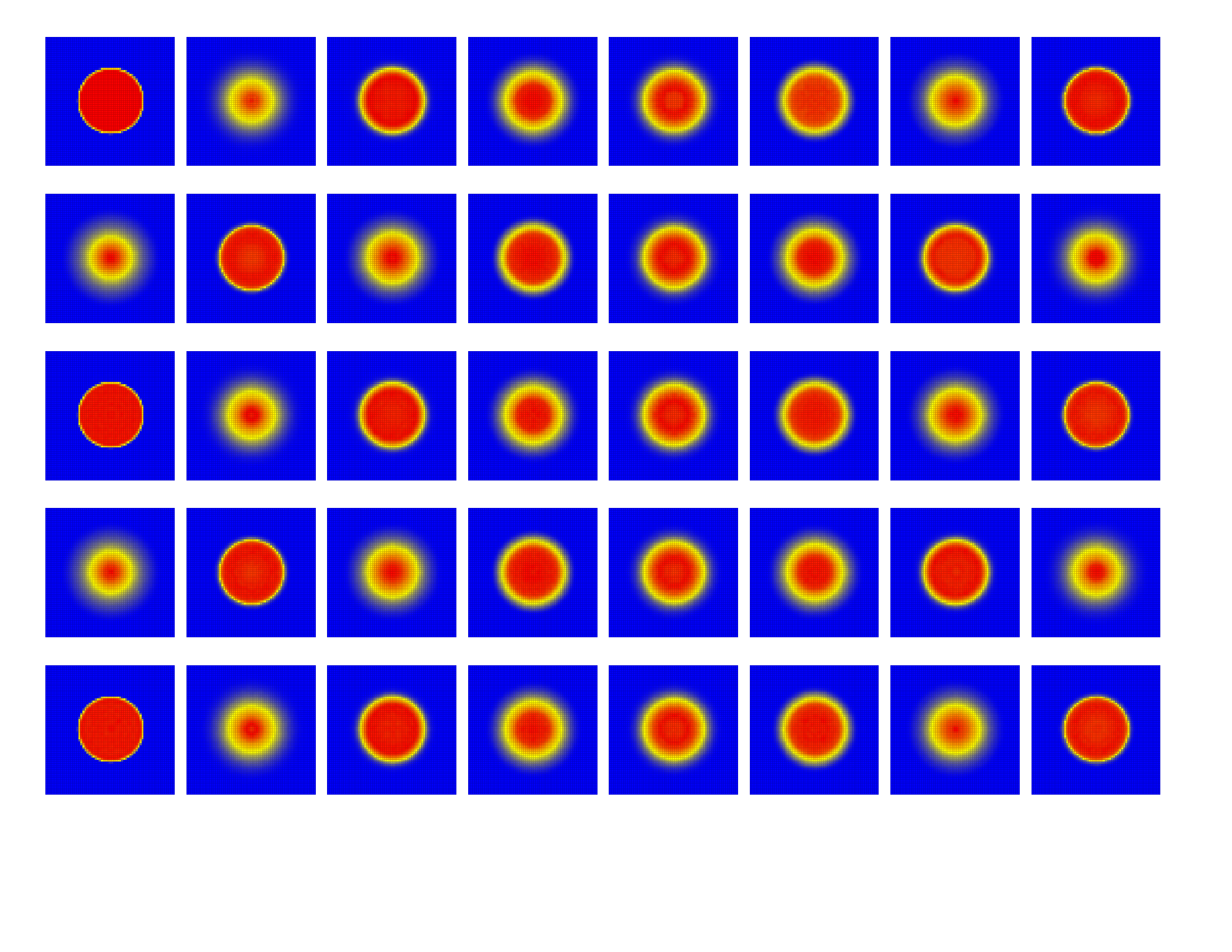}
\caption{Time evolution of the density $|\psi(\bs r,t)|^2$ calculated from the solution of the Gross--Pitaevskii equation for an initial disk  box potential (see text for details). Each image correspond to time $t=jT/8$, with $j=0,\ldots,39$. The evolution is approximately periodic with period $2T$. }
\label{fig:evol_disk}
\end{figure*}

We show in figures \ref{fig:evol_triangle}-\ref{fig:evol_disk} the time evolution of the density distribution, calculated using the numerical solution of the time-dependent Gross--Pitaevskii equation. The initial distribution corresponds to the ground state in a box-like potential. The box has the shape of a regular polygon (triangle, square, pentagon, hexagon) or a disk. It is centered on the calculation grid, and its summits are located on a circle of diameter $L/2$, where $L$ is the total size of the grid. As in the experiment, the box confinement is removed at time $t=0$ and the wave function then evolves in an isotropic harmonic confinement with frequency $\omega$. The frequency $\omega$ of the harmonic potential is chosen such that $E_{\rm p}(0)\approx E_{\rm tot}/2$.

Here the calculations are performed on a grid $512\times 512$, with an initial state in the Thomas--Fermi regime. The healing length is $\xi \approx 1.4\,\ell$ for the  triangular case and $\xi\approx 2\,\ell$ for the other cases, where $\ell$ is the grid spacing.  Each panel contains 40 images, which display (from left to right and from top to bottom) the density distribution at time $t=jT/8$, with $j=0,\ldots,39$. 

\begin{figure*}[t]
  \includegraphics{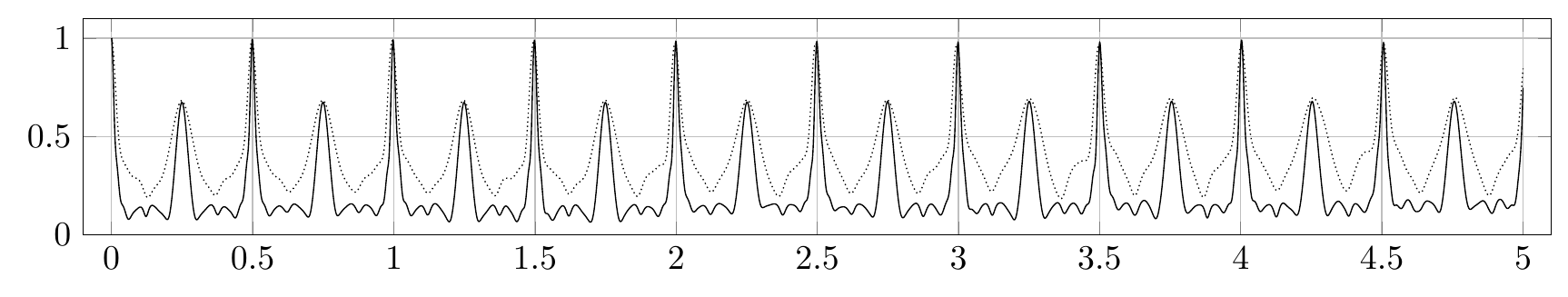}\\
  \includegraphics{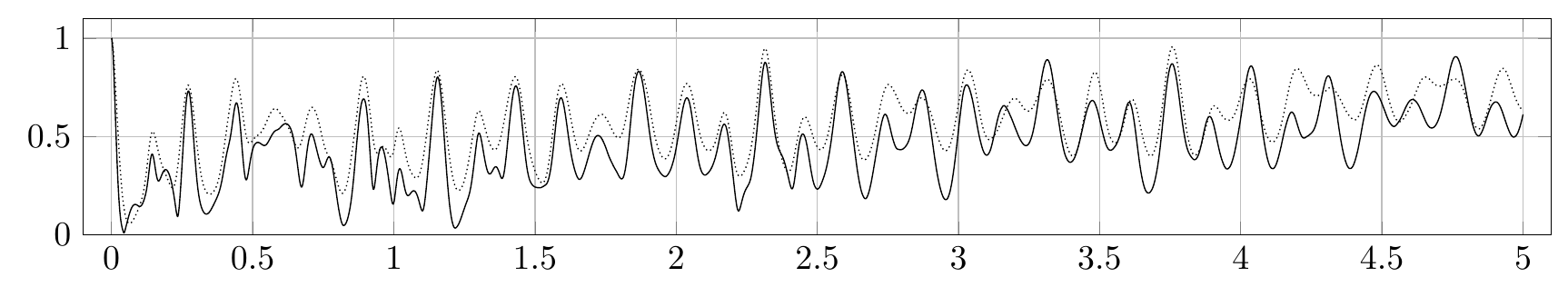}\\
  \includegraphics{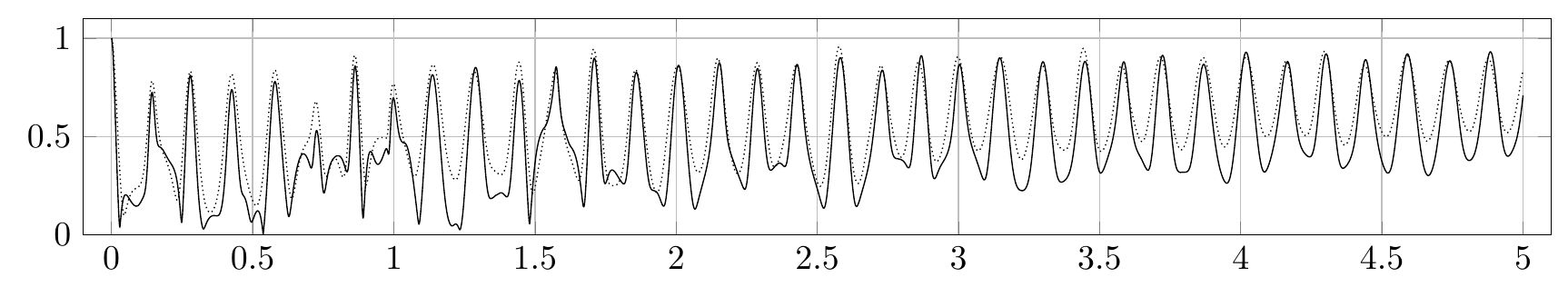}\\
  \includegraphics{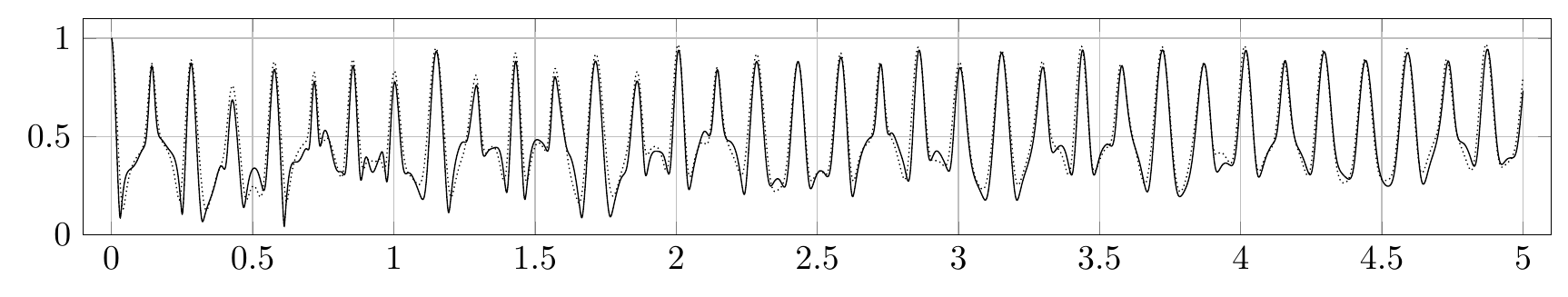} \\
  \includegraphics{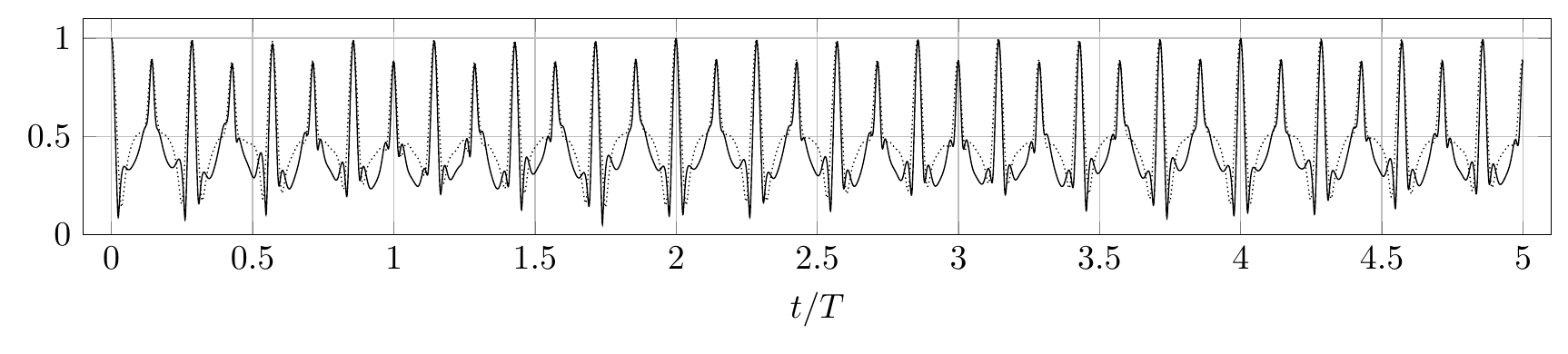}   
  \caption{Evolution of the overlap $|\langle \psi(0)|\psi(t)\rangle|$ as a function of $t/T$. From top to bottom: triangle, square, pentagon, hexagon, disk. Continuous line: calculation for a grid $512\times512$ and the same ratio $\xi/\ell$ as for the gallery of images \ref{fig:evol_triangle}--\ref{fig:evol_disk}. Dotted line: Evolution calculated for a smaller grid ($256\times 256$) and $\xi/\ell\approx 1.4$.}
\label{fig:ovlps}  
\end{figure*}

We show in figure \ref{fig:ovlps} the evolution of the overlap $|\langle \psi(0)|\psi(t)\rangle|$ for the set of data of Figs. \ref{fig:evol_triangle}--\ref{fig:evol_disk}  (grid $512\times 512$), together with the evolution of the same quantity calculated on a smaller grid ($256\times 256$, dotted line). The periodicity $T/2$ (resp. $2T$) appears clearly for an initially triangular (resp. disk-shaped) cloud. For these two cases, the local maxima of the overlap get closer to unity as the grid size increases, and the evolution of the overlap is approximately symmetric around the times $t=jT/4$ for the triangles ($t=jT$ for the disks), as expected from time reversal symmetry arguments for a periodic evolution (see main text). On the opposite, no such hint for a periodic behavior shows up on this time interval for a cloud whose initial shape is a uniform square, pentagon or hexagon.

\end{document}